\documentclass[a4paper,fleqn,usenatbib,useAMS]{mnras}

\usepackage{graphicx}	
\usepackage{amsmath}	
\usepackage{amssymb}	
\usepackage{multicol}        
\usepackage{bm}		
\usepackage{pdflscape}	

\newcommand{\kms}{\,km\,s$^{-1}$} 
\newcommand{\mt}[1]{\mathrm{#1}}

\usepackage[T1]{fontenc}
\usepackage{ae,aecompl}

\usepackage{newtxtext,newtxmath}

\title[Stellar escapers from M67]{\textit{Stellar escapers from M67 can reach solar-like Galactic orbits}}

\author[T. G. J\o rgensen]{Timmi G. J\o rgensen$^{1}$\thanks{Contact e-mail: \href{mailto:timmi@astro.lu.se}{timmi@astro.lu.se}}, Ross P. Church$^{1}$
\\
$^{1}$Department of Astronomy and Theoretical Physics, Lund Observatory, Box 43 SE-221 00 Lund, Sweden}

\date{Accepted XXX. Received YYY; in original form ZZZ}

\pubyear{2019}

\begin{document}
\label{firstpage}
\pagerange{\pageref{firstpage}--\pageref{lastpage}}
\maketitle

\begin{abstract}
We investigate the possibility that the Sun could have been born in M67 by carrying out $N$-body simulations of M67-like clusters in a time-varying Galactic environment, and following the galactic orbits of stars that escape from them. We find that model clusters that occupy similar orbits to M67 today can be divided up into three groups. Hot clusters are born with a high initial $z$-velocity, depleted clusters are born on cold orbits but are destroyed by GMC encounters in the Galactic disc, and scattered clusters are born on cold orbits and survive with more than 1000 stars at an age of 4.6 Gyr. We find that all cluster models in all three cluster groups have stellar escapers that are kinematicaly similar to the Sun. Hot clusters having the lowest such fraction $f_{\odot} = 0.06$ \%, whilst depleted clusters have the highest fraction, $f_{\odot} = 6.61$ \%. We calculate that clusters that are destroyed in the Galactic disc have a specific frequency of escapers that end up on solar-like orbits that is $\sim$ 2 times that of escapers from clusters that survive their journey.     

\end{abstract}




\begingroup
\let\clearpage\relax
\endgroup
\newpage

\section{Introduction}
M67 is an old, metal-rich open stellar cluster located at a distance of around 800 - 900 pc from the Sun. It is at a height of $z$ $\sim$ 400 - 450 pc (\citet{Sarajedini2009}, \citet{VandenBerg2004}, \citet{Friel1995}) above the Galactic Plane which has allowed the cluster to survive for much longer than open clusters usually do, with an estimated age of 3.5 - 4.8 Gyr \citep{Yadav2008}. 

The links between the Sun and M67 are numerous. M67's estimated age is comparable to the age of the Sun, together with the cluster having a solar-similar chemical composition with [Fe/H] = 0.00 $\pm$ 0.06 \citep{Heiter2014}. A dominant fraction of stars analysed in M67 by \citet{Onehag2014} shows abundance patterns that are in better agreement with the Sun than solar twins in the Solar neighbourhood.

There are several indications from our Solar system that suggest a Solar birth cluster of substantial size such as M67 is needed. Meteoritic evidence of short-lived radioactive isotopes implies that the early Solar System has been enriched by an external source. A popular external source for enrichment is a supernova explosion with an preferred initial starting mass of $\sim$ 25 $M_{\odot}$ which has to be within a distance of $\sim$ 0.1 - 0.3 pc in order to provide enough enrichment for the Solar nebula \citep{Adams2010}. From cluster enrichment simulations with a total number of 2100 stars \citet{Parker2014} found that this enrichment process is able to enrich $\sim 1$ \% of G-dwarfs that have always been single stars and that this process is quite stochastic. 

The outer objects of the Solar System also provide clues to the Sun's birth enviroment. With an large eccentricity of $\sim$ 0.84, Sedna has an unusual orbit compared to other solar system objects. Numerical studies \citep{Morbidelli2004} have shown that Sedna's orbital elements can be explained by a close encounter with a passing star where an impact parameter of $\sim$ 400 AU is needed. 

The connection between M67's position high above the Galactic plane and the Sun's position has been addressed by \citet{Pichardo2012} who investigated their dynamical link by integrating the current positions of M67 backwards in time in a galactic model which included spiral arms and a Galactic bar. They argued that the Sun could not be born in M67, since the outer part of the solar system would have been destroyed as the Sun was ejected from the cluster onto the current solar orbit. This conclusion is based on the assumption that M67 was born on its current orbit. However, as shown by \citet{Gustafsson2016}, if M67 was born in an orbit close to the Galactic plane and was then subsequently scattered to high altitudes by encounters with giant molecular clouds (GMCs), the kick needed for the Sun to escape M67 would not need to be so strong as to destroy the outer Solar System.    

In this paper we investigate the orbits of stellar cluster escapers and the possibility that the Sun has been born in M67 by using a Milky Way model and $N$-body simulations of M67. We test the hypothesis that the Sun was tidally stripped away from M67 by interactions with the spiral arms and GMCs of the Milky Way as the cluster departed from its birth place in the Galactic plane to a height of $z$ $\sim$ 400 - 450 pc (\citet{Sarajedini2009}, \citet{VandenBerg2004}, \citet{Friel1995}). The tidal stripping is caused by the galactic potential, which in our model contains both a static component and time-varying components that represent the spiral arms, Galactic bar and GMCs. Interactions with this time-varying force field can scatter clusters such as M67 up to high $|z|$ as shown by \citet{Gustafsson2016}. 

The simulations we preform can be divided into three steps. \textbf{Step 1:} We follow test particles, representing clusters, in the Milky Way model for 4.6 Gyrs and identify those that have similiar orbits to the current orbit of M67. We call these M67 candidates. \textbf{Step 2:} The evolution of each M67 candidate is computed in a $N$-body simulation where the effects of the galactic tidal field and encounters of the cluster with the GMCs are taken into account. \textbf{Step 3:} We follow the escaped stars from the $N$-body simulations in the Milky Way model til the present day and analyse their orbits.  

The paper is organized as follows: In Section \ref{sec:MW} we present the galactic model we use to represent the Milky Way and follow galactic orbits. In Section \ref{sec:C_surv} we define M67-like candidates in terms of their orbital parameters and survivability based on their GMC encounters. Section \ref{sec:Nbody} describes the parameters of the $N$-body model that is used to represent an initial model of M67. We present the tidal tensor which represents the galactic tidal field and explain how GMC encounters are incorporated. We show our results in Section \ref{sec:Results} and discuss the different M67 model histories in terms of their orbital journey to the present day position of M67 and what orbits their stellar escapers end up on. We further discuss the results in Section \ref{sec:Discussion} and give our conclusions in Section \ref{sec:Conclusions}.

\section{Milky Way Model}
\label{sec:MW}
The Milky Way model follows \citet{Gustafsson2016} and consists of a axisymmetric potential \citep{Binney2012}, two spiral arms \citep{Pichardo2003}, a Galactic bar \citep{Pichardo2012} and a distribution of GMCs along the spiral arms \citep{Gustafsson2016}.

The axisymmetric galactic potential in our model is based on Potential I of \citet{Binney2012} which consists of a thin and thick stellar disc, a gas disc, and a stellar and dark matter spheroid representing the Bulge and the Halo. From the potential we removed the stellar spheroid representing the Bulge and replaced it with our model of the Galactic bar (see below). This gives a circular speed of $\sim$ 220 \kms \, at $R_0=8$ kpc.
\subsection{Spiral arms}
The spiral pattern consists of two spiral arms with a pitch angle of 15.5$^{\circ}$ described by \citet{Pichardo2003}. Each arm is represented by 100 inhomogenous oblate spheroids with semi-major and minor axes of 1000 pc and 500 pc, respectively. The spiral pattern starts at the radius where the Galactic bar ends at $R_s$ = 3.13 kpc with a distance between each spheroid of 500 pc. The spheroid density follows a linear density law
\begin{equation}
\rho(a) = \rho_0 ( 1 - a),
\label{eq:densitylaw}
\end{equation}
where $a = (x^2 + y^2 + z^2/c_0^2)^{1/2}$ and $c_0 = 500$ pc. This density law gives each spheroid a zero density at its boundary where $\rho_0$ is the central density. The central density for each spheroid has a exponential decline of 
\begin{equation}
\rho_0(R) = \rho_{02} \, e^{(-(R - R_s)/R_L)},
\label{eq:rho}
\end{equation}
with a radial scale height, $R_L$, of 3.9 kpc. $\rho_{02}$ is given in \citet{Pichardo2003} by
\begin{equation}
\rho_{02} = \frac{ 3 M_s}{2 \pi a_0^2 c_0 \sum_{j=1}^{N} e^{-(R_j - R_s)/R_L}},
\label{eq:rho02}
\end{equation} 
where $M_s$ is the total mass in the spiral arms, $R_j$ is the Galactocentric distance of the spheroid's centers and the sum is over one spiral arm. In our model we use a total spiral mass of $4 \times 10^9$ $M_{\odot}$ and a constant pattern speed of 24 \kms \, kpc$^{-1}$.

The force from each spheroid was pre-calculated in a grid as a function of radius $r$ and angle $\beta$ in the $x-z$ plane of the spheriod and used in a bicubic interpolation for the force calculation for each oblate spheroid representing the spiral arms. Due to symmetry $\beta$ was defined between $0^{\circ}$ and $90^{\circ}$ with a step size of $\Delta \beta = 0.5 ^{\circ}$. The radius $r$ was defined between 0 and 20 kpc with a step size of 0.1 kpc, except for the inner 2 kpc where the step size was 0.01 kpc in order to get higher resolution for a more precise bicubic interpolation. 
\subsection{Galactic bar}
The Galactic bar is represented by a prolate inhomogenous spheroid \citep{Pichardo2003, Pichardo2012} with a linear density law similar to Eq. \ref{eq:densitylaw} with zero density at the boundary. The semi-major and minor axes of the Galactic bar is $a_0 = 3.13$ kpc and $c_0 = 1.0$ kpc which gives $a = (x^2/a_0^2 + y^2 + z^2)^{1/2}$. The force exerted by the Galactic bar was pre-calculated in a grid as a function of $r$ and angle $\beta$. The radius in the grid was defined between 0 to 12 kpc with a step size of 0.1 kpc. The range of $\beta$ was between $0^{\circ}$ and $90^{\circ}$ with a step size of $\Delta \beta = 1.0 ^{\circ}$. The numerical integration which was used to produce the pre-calculated grid was done in a similar fashion as for the spiral arms. The total mass of the Galactic bar was set to $1.6 \times 10^{10}$ $M_{\odot}$ with a constant pattern speed of 55 \kms \, kpc$^{-1}$.
\subsection{Giant Molecular Clouds}
\begin{figure}
 \includegraphics[width=\columnwidth]{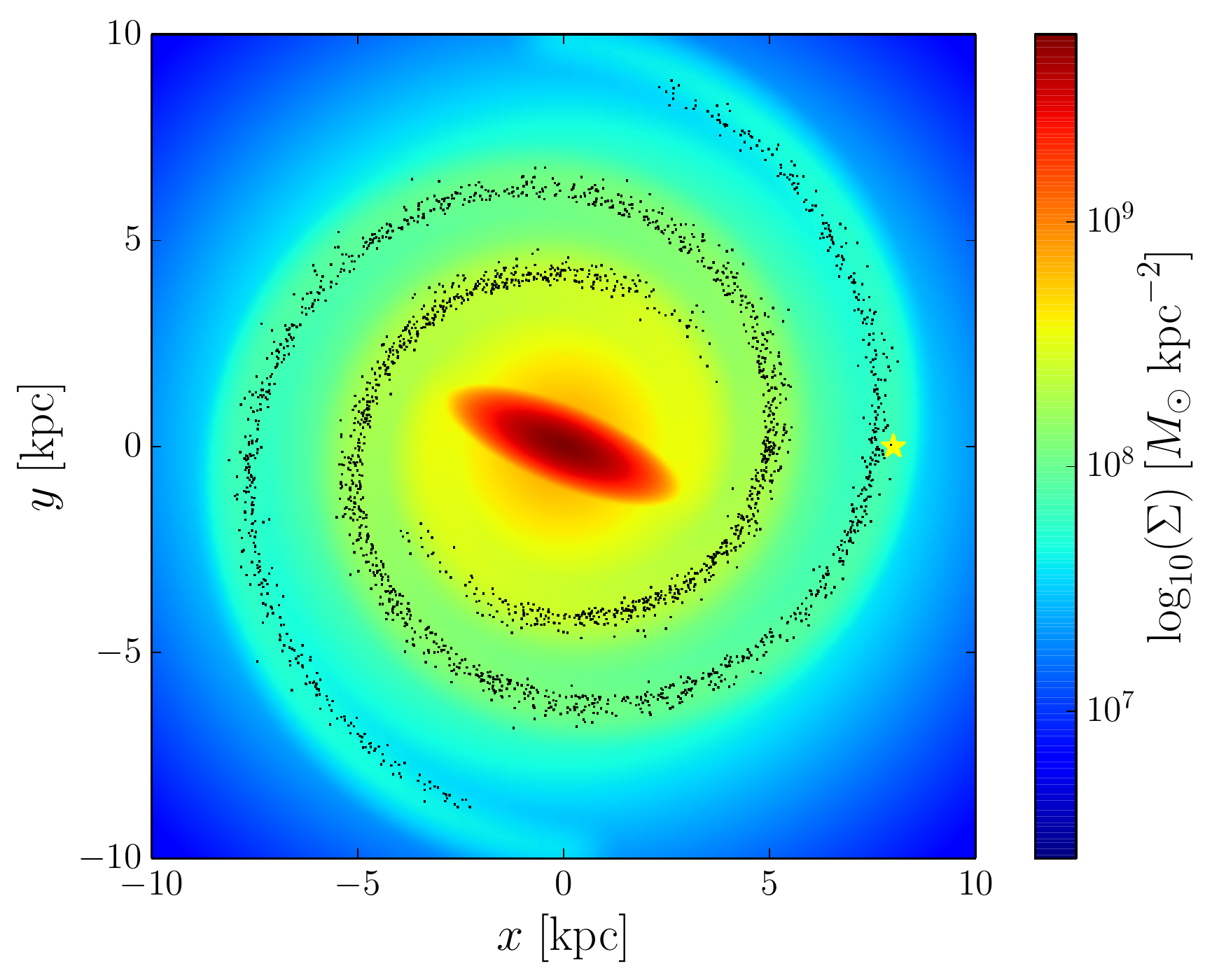}
 \caption{Surface density profile of the Milky Way model at present time. The black dots represent the GMCs currently located in the model while the yellow star represents the location of the Sun. The spiral arms and Galactic bar rotate with a constant pattern speed of 24 and 55 \kms \, kpc$^{-1}$, respectively. They remain present at all times throughout the simulations while the GMCs are transient with a lifetime of 40 Myr.}
 \label{fig:MW_profile}
\end{figure}

The implementation of GMCs follows \citet{Gustafsson2016} where the maximum mass of a GMC, $M_i$, was selected according to the distribution function
\begin{equation}
\frac{\mt{d}N}{\mt{d}M_i} \propto M_i^{-1.8} \, \, \, \, \, \, \, \, \, \, \, \, \, \, \, \, \, \, \mathrm{with} \, \, \, \, \, \, \, \, \, \, \, \, \, \, 5.0 \le \mt{log}_{10}(M_i/M_{\odot}) \le 7.0,
\end{equation}  
based on the observations and simulations of \citet{Williams1997} and \citet{Hopkins2012}, respectively. We give each GMC a lifetime of 40 Myr corresponding to a few free fall times. During its lifetime the GMC will increase its mass to a value of $M_i$ over 20 Myr and then decrease back to zero over the next 20 Myr with a parabolic mass evolution of
\begin{equation}
M_t = \left [ -0.25 \cdot \left (\frac{t-t_0}{10^7} \right )^2 + \frac{t-t_0}{10^7}  \right ] \cdot M_i,
\label{eq:M_t}
\end{equation}
where $t_0$ is the birth time of the GMC and $t$ is the current time, both in years. The initial birth times for the first GMCs were chosen randomly between the time -40 and 0 Myr. When the life of one GMC ended a new is born such that the number of GMCs in our simulations is always constant. We adopt the total Galactic molecular mass of $1.0 \times 10^9$ $M_{\odot}$ from \citet{Williams1997} resulting in a constant number of 2500 GMCs. Each GMC is represented by a \citet{Plummer1911} sphere with a core radius of
\begin{equation}
R_c = 20 \cdot \left ( \frac{M_t}{5 \times 10^5 M_{\odot}} \right )^{1/2} \mt{pc}.
\label{eq:R_c}
\end{equation} 
Newborn GMCs are distributed within a distance of $\pm$ 50 pc from the median line of the spiral arms in the Galactic plane, at Galactocentric distances 4 kpc $\leq  R_0 \leq$ 9kpc, with a number density decreasing exponentially along the arms similar to Eq. \ref{eq:rho} in order to match the density decrease of the arms. Each GMC initially moves with the local circular rotation velocity plus a velocity scatter which has a Gaussian distribution of $\sigma = 7$ \kms \, for each of the three velocity components. 
The total surface density profile for the Milky Way model can be seen in Figure \ref{fig:MW_profile}. The position of the Sun is represented by the yellow star and the black dots are the GMCs at a time of 4600 Myr into the simulation, corresponding with our present time.   
\subsection{Matching Milky Way model with observations}
\label{sec:MW_obs}  
The test particles representing our clusters were randomly placed in the Galactic disc with Galactocentric distances, 4 kpc < $R$ < 10 kpc. Like the GMCs, the particles were given a gaussian velocity distribution of 7 \kms \, in each of the three velocity directions, together with an initial local circular rotation speed. The particles were followed for 4.6 Gyr with a total number of test particles of 25000 divided into five independent runs. Each run was done in a fixed reference frame with a rotating Galactic bar and spiral arms. A constant timestep of 0.1 Myr was used for all Milky Way model runs. 

\begin{figure}
 \includegraphics[width=\columnwidth]{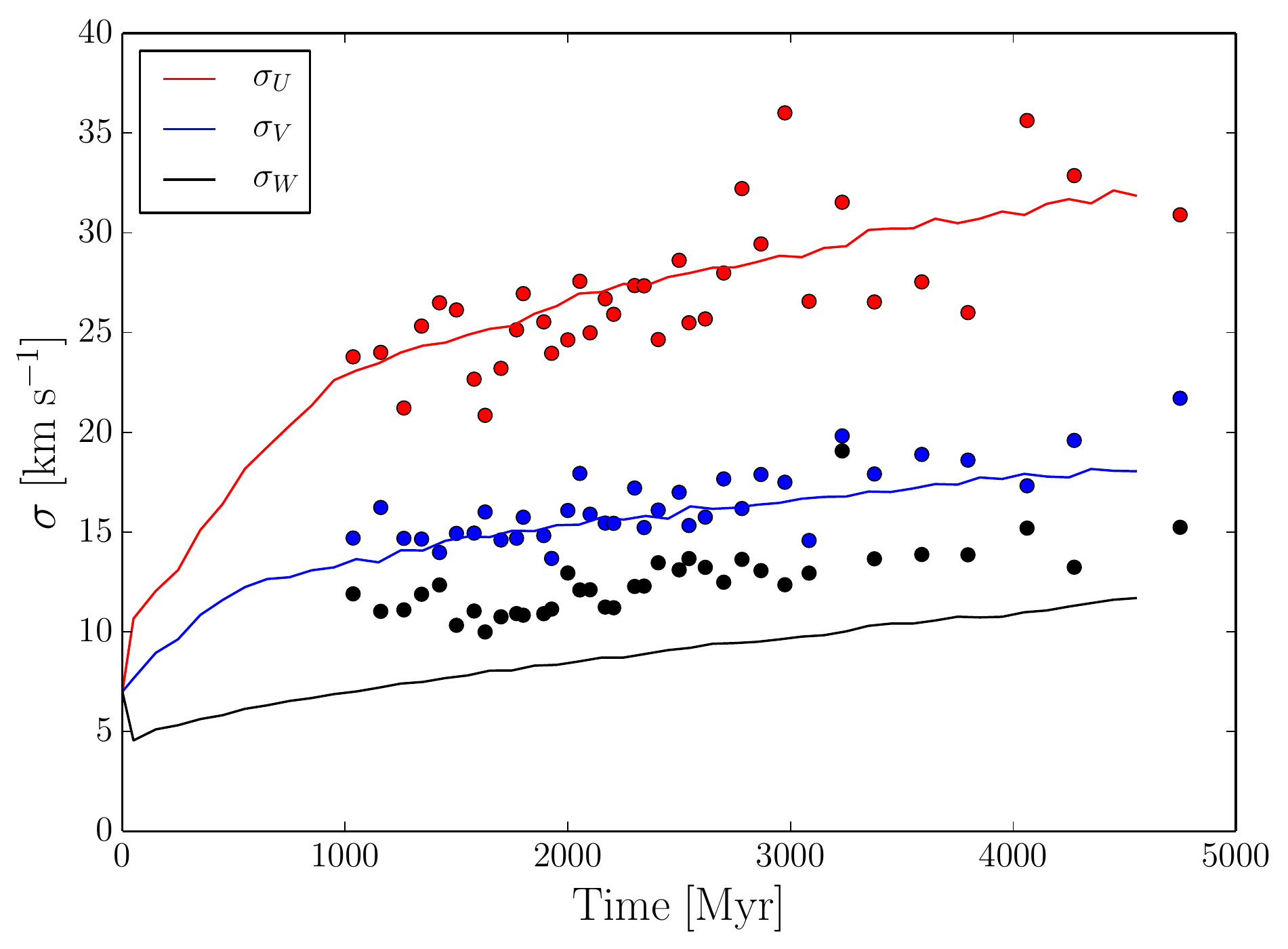}
 \caption{The velocity dispersion of stars in the three galactic rectangular coordinates $U,V$ and $W$ as a function of time for our galactic model (lines). The velocity dispersions are calculated for stars within 300 pc of $R = 8$ kpc from the five Milky Way simulations, with a binned $\Delta t$ of 100 Myr. The points represent the stellar observational results of \citet{Holmberg2009} which used the Hipparcos parallaxes to improve the accuracy of the Geneva-Copenhagen Survey \citep{Nordstrom2004} for nearby solar-type stars. Each data point is a bin containing 150 stars, where the time is given as the average age of the stars.}
 \label{fig:Dispersions_time}
\end{figure}

Our model was able to reproduce the observed velocity dispersion of solar-type stars within 300 pc of the Sun by \citet{Holmberg2009} as a function of time seen in Figure \ref{fig:Dispersions_time}. The test particles used for calculating the velocity dispersion were taken from within 300 pc of a Galactocentric radius of 8 kpc for all azimuthal angles in order to get better number statistics. We see that the evolution of $\sigma_{U}$ and $\sigma_{V}$ follows the observations quite well. The evolution of $\sigma_{W}$ does not follow the observations and is displaced by a few \kms \, which suggest that our Milky Way model is biased against test particles going onto high altitudes. For the purpose of our investigation, this means that scattering test particles into M67-orbits in our Milky Way model will be more difficult and thus will most likely underestimate the number of M67 candidates in our model.

\subsection{Solar-like orbit}
\begin{figure*}
\centering
\begin{tabular}{|c|c|}
 \includegraphics[width=\columnwidth]{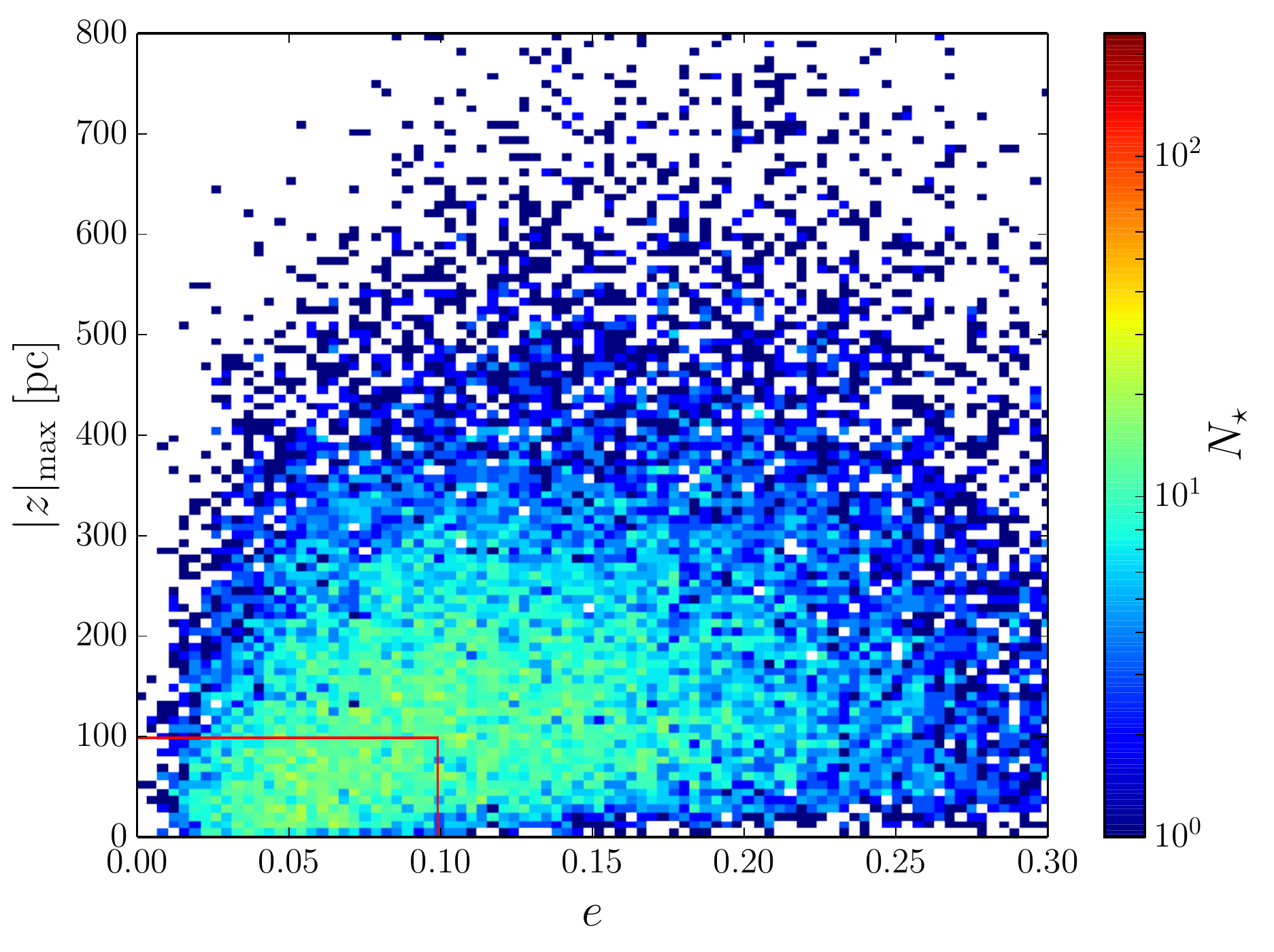} & \includegraphics[width=\columnwidth]{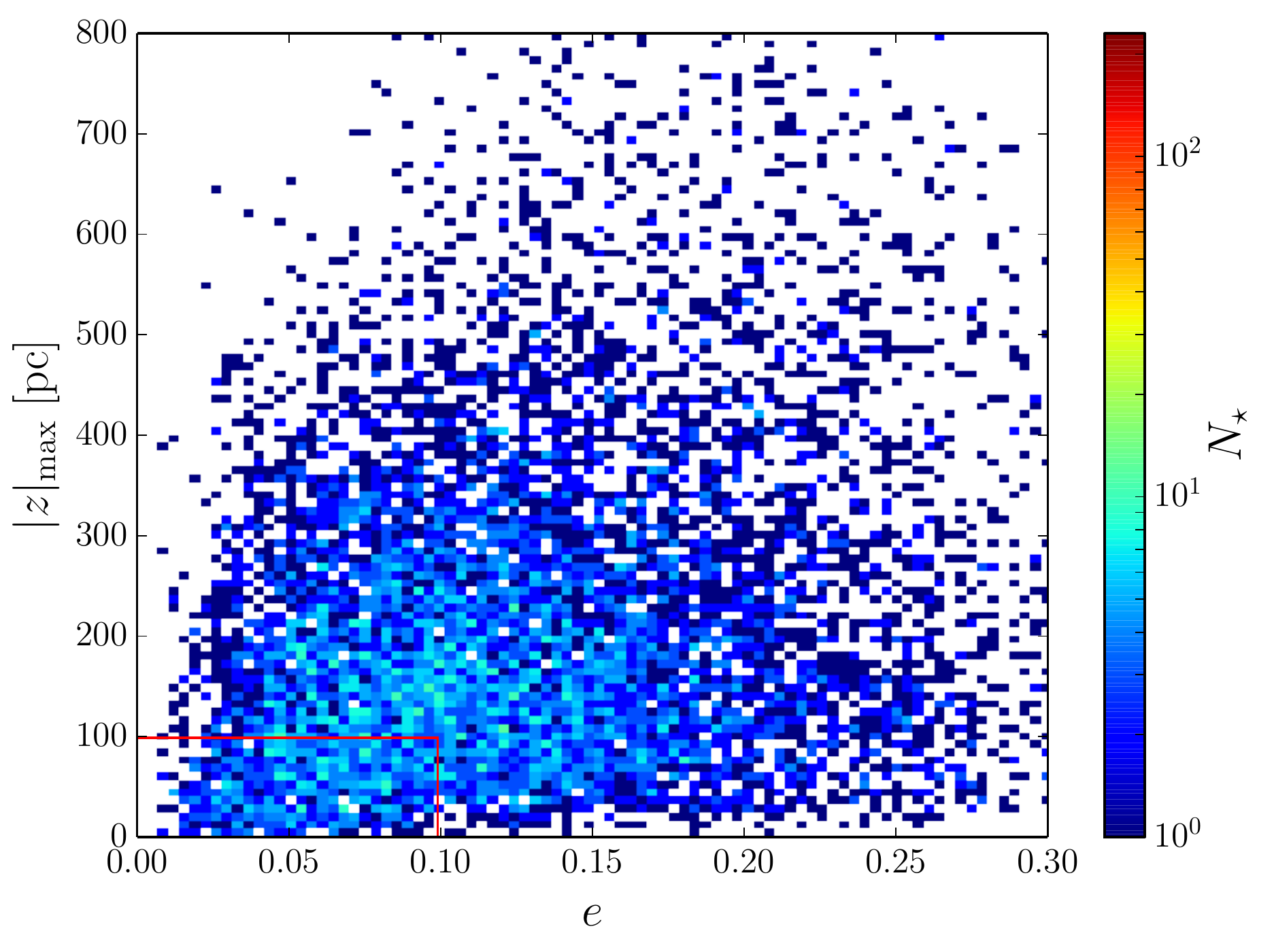} \\		 	
\end{tabular}
\caption{\textit{Left:} The final positions of the 25,000 test particles in the $e$-$|z|_{\rm{max}}$ plane from the Milky Way model simulations after a time of 4.6 Gyr. \textit{Right:} The subset of the test particles to the \textit{left} that have a guiding radius between 7 and 9 kpc resulting in a total of 8578 test particles. The red rectangle marks the region where the stars are defined to be solar-like.}
\label{fig:test_particles}
\end{figure*}

In order to compare the orbits of escaped stars with that of the Sun we first need to define a Solar-like orbit. We use three parameters to classify orbits in the Milky Way model: the eccentricity $e$, the maximum height above the Galactic plane $|z|_{\mathrm{max}}$ and the guiding radius $R_g$. The eccentricity was calculated as $e = (r_a - r_p)/(r_a + r_p)$, where $r_a$ and $r_p$ are the apoapsis and periapsis of a single orbit, which we define as a full rotation of 360$^{\circ}$ in the Milky Way disc. $|z|_{\rm{max}}$ is defined as the maximum $|z|$ of a test particle during a full orbit. The guiding radius was defined as $R_g = L_z/v_c$, where $L_z$ is the $z$-component of the angular momentum and $v_c$ is the local circular velocity at the test particle's current position. To construct the orbit we used the local standard of rest and the current $z$-position of the Sun. We adopted the uncertainty of the Galactocentric distance $\delta R$ from \citet{Brunthaler2011} with a Galactocentric distance of 8 kpc as defined in our Milky Way model, the parameters of which can be seen in Table \ref{table:MW}. We integrated the orbit of the Sun 300 Myr back in time to obtain $e$, $|z|_{\mathrm{max}}$ and $R_g$. No GMCs were used for the backwards integration since this could result in an unrealistic higher spread of the three parameters. The backwards integration of the Sun was done for 5000 test particles with initial values sampled according to a Gaussian distribution of the values in Table \ref{table:MW}. From this we obtained $e_{\odot} = 0.1$ and $|z|_{\mathrm{max},\odot} = 98.7$. In order for cluster escapers to end up on solar-like orbits, they need to be dynamically colder than the current values of $e_{\odot}$ and $|z|_{\mathrm{max},\odot}$ and not necessarily match these values. We do not require the guiding radius to match the Solar value as a criteria for escapers to be classified as solar-like. This is because $R_g$ is mainly modified by churning of orbits by the spiral arms  and not scattering off GMCs. The spiral arms are transient in the Milky Way but static in our model, which means that the position in terms of guiding radius of our escapers could be quite different depending on the spiral arm evolution of the Milky Way during the last 4.6 Gyrs. In regards to how normal it is to be solar-like in our model we compare the final position of our 25000 test particles to our definition of what we classify as a solar-like orbit. To the \textit{left} in Figure \ref{fig:test_particles} the position of all 25000 particles at a time of 4.6 Gyr are shown, with the red rectangle defining the area where test particles are solar-like. To the \textit{right} in Figure \ref{fig:test_particles} we only see the 8578 particles that have a guiding radius between 7 and 9 kpc. The fractions of particles that have solar-like orbits at the end of the simulations is quite high, 13 \% in the full sample and 11 \% in the subset with 7 $\le R_g \le$  9 kpc. We would therefore expect that a high fraction of escapers, from clusters that are destroyed in their early life in the Galactic disc, will be on solar-like orbits at a time of 4.6 Gyr.

\section{Cluster survival and M67-like orbits}
\label{sec:C_surv}
\begin{figure}
 \includegraphics[width=\columnwidth]{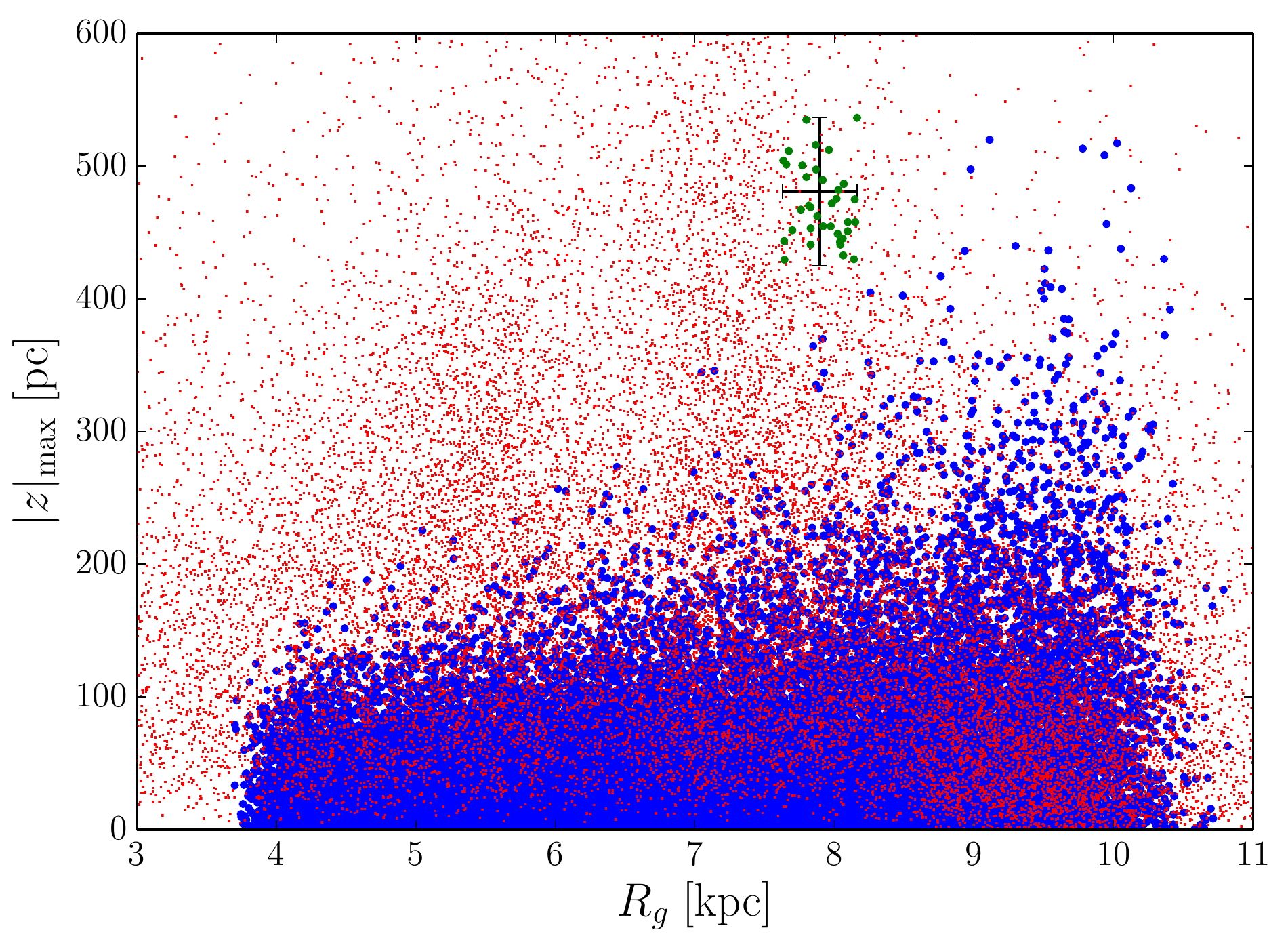}
 \caption{Here we see the test particles of the Milky Way model in the $e$-$|z|_{\rm{max}}$ plane. The blue points represents the cluster particles after they have completed their first full orbit in the Milky Way simulation at a time of $\sim$ 200 Myr. The red points are the final positions of the cluster particles after 4600 Myr. The errorbar denotes the current location of M67 and the green points are the cluster particles which have an identical orbit to M67 in $R_g$, $|z|_{\mathrm{max}}$ and $e$. All particles are born with the same initial velocity dispersion of 7 \kms \, in the $z$-direction, therefore particles at high $R_g$ have a higher average $|z|_{\rm{max}}$, since the gravitational potential is weaker at larger $R_g$.}
 \label{fig:initial_vs_final}
\end{figure}

We define M67-like orbits in the same way as we define the orbital parameters for the Sun. We constructed the orbit using the radial velocity, proper motion and position of M67 seen in table \ref{table:MW}. With the absence of GMCs we integrated the orbit of M67 300 Myr back in time to get $e$, $|z|_{\mathrm{max}}$ and $R_g$. We found the orbital parameters of M67 to be $R_g = 7.89\pm0.27$ kpc, $|z_{max}| = 480\pm56$ pc and $e = 0.12\pm0.02$. Out of the 25000 test particles in our Milky Way model, 37 of them had similar orbital parameters as M67 to within one standard deviation and can be seen as green points in Figure \ref{fig:initial_vs_final} where the location of the initial cluster test particles can be seen in blue. The red points are the position of the cluster particles after 4600 Myr. 

GMCs can help scatter clusters to high galactic orbits, but they can also destroy the clusters via tidal forces if the GMC comes too close. We therefore had to evaluate which of our 37 cluster particles would survive their journey to the location of M67 over a time span of 4.6 Gyr. We consider encounters between a GMC and a cluster particle to be close when the closest separation $r_{\mathrm{sep}} \le d$ where 
\begin{equation}
d = 100 \cdot \left ( \frac{M_i}{10^6 M_{\odot}} \right )^{1/2} \, \mathrm{pc}.
\label{eq:GMC_reg}
\end{equation} 
\begin{table}
 \caption{Observational data and their transformation into Milky Way model parameters for the Sun and M67. References are: (a) \citet{Bellini2010}, (b) \citet{Schonrich2010}, (c) \citet{Reed2006} and (d) \citet{Brunthaler2011}.}
 \label{table:MW}
 \begin{tabular*}{0.50\textwidth}{@{\extracolsep{\fill}} lcc}
  Observational parameter & Value & Reference \\
  \hline
  \hline
   & M67 & \\
  \hline
  $\mu_{\alpha}$ cos($\delta$) & -9.6$\pm$1.1 [mas yr$^{-1}$]  & a \\
  $\mu_{\delta}$ & -3.7$\pm$0.8 [mas yr$^{-1}$] & a \\
  Radial velocity & 33.78$\pm$0.18 [\kms] & a \\ 
  Distance &  815$\pm$81.5 [pc] & a\\ 
  \hline
  & Sun & \\
  \hline
  $U_{\odot}$ & -11.1$\pm$1.2 [\kms] & b\\
  $V_{\odot}$ & 12.24$\pm$2.1 [\kms]&  b\\
  $W_{\odot}$ & 7.25$\pm$0.6 [\kms]&  b\\
  $z_{\odot}$ & 20$\pm$5 [pc] & c \\
  $\delta R$ & 0.23 [kpc] & d\\	 
 \end{tabular*}
 \begin{tabular*}{0.50\textwidth}{@{\extracolsep{\fill}} ccc}
  \hline
  \hline
  & Milky Way model & \\
  \hline
 \end{tabular*}
 \begin{tabular*}{0.50\textwidth}{@{\extracolsep{\fill}} lccc}
    & $e$ & $|z|_{\mathrm{max}}$ & $R_g$ \\
    &  & $[\mathrm{pc}]$ & $[\mathrm{kpc}]$ \\	
  \hline
  Sun & 0.10$\pm$0.01 & 98.7$\pm$9.2 & 8.4$\pm$0.3 \\
  M67 & 0.12$\pm$0.02 & 480$\pm$56 & 7.9$\pm$0.3 \\
  \hline
  \end{tabular*}
\end{table}
Following Eq. (19) from \citet{Gustafsson2016}, the fractional change in kinetic energy in a cluster as it encounters a GMC can be approximated as 
\begin{equation}
\delta_E= \frac{8 G M^2 r_h^3}{3 m b^4 V^2},
\label{eq:d_E}
\end{equation} 
where $G$ is the gravitational constant, $M$ is the mass of the GMC, $r_h$ is the half-mass radius of the cluster, $m$ is the mass of the cluster, $V$ is the relative speed between the GMC and cluster and $b$ is the impact parameter. If the cumulative $\delta_E > 1$ for all the GMC encounters, the cluster is tidally destroyed. For the half-mass radius of M67 we chose, in accordance with previous investigations of M67 in \citet{Hurley2005} a value of 4 pc and 36000 stars corresponding to a mass of $\sim2.2 \times 10^4$ $M_{\odot}$. The values of the half-mass radius and mass will change as the cluster evolves but remained constant in Eq. \ref{eq:d_E} as a conservative estimate of the cluster survivability. Out of the 37 M67-like clusters we estimate that 16 will survive their GMC encounters. We classify these 16 clusters as M67 candidates and further investigate their histories.   

\section{$N$-body simulations}
\label{sec:Nbody}
In order to obtain realistic escape histories of our 16 M67 candidates we carried out $N$-body simulations for each cluster. The GMC encounter histories and orbits in the galactic field of our Milky Way model will shape the evolution of each M67 candidate differently. To follow the galactic tidal enviroment of each cluster we used the $N$-body code \textsc{nbody6tt} \citep{Renaud2011}. \textsc{nbody6tt} is a modified version of \textsc{nbody6} \citep{Aarseth2003} where a tidal tensor can be implemented to described the tidal forces the cluster experiences during its orbit.
\subsection{Initial parameters of M67}
For the initial parameters of M67 we adopt initial values similar to those used by \citet{Hurley2005}, with 36,000 stars and a half-mass radius of 4 pc. The stars were distributed in mass according to a \citet{Kroupa2001} IMF in a range of 0.1 - 50 $M_{\odot}$ and spatially distributed according to a \citet{Plummer1911} distribution. Stellar evolution was turned on with a solar metallicity ($Z = 0.02$).
\subsection{Tidal tensor}
The tidal tensor is defined in \citet{Renaud2011} as 
\begin{equation}
T_{ij} = -\frac{\partial^2 \Phi}{\partial x_i \partial x_j},
\label{eq:tt}
\end{equation}
where $\Phi$ is the gravitational potential of the galactic enviroment at the cluster's position. The tidal tensor was constructed by following each M67 candidate through the Milky Way model. We calculated $T_{ij}$ by numerical differentiation of the force for each time step within a cube of size $2\delta$ centered on the cluster particle. The value of $\delta$ was chosen to be 15 pc, corresponding to a few half-mass radii of the cluster, such that the tidal tensor would be able to resolve the fine structures of the gravitational potential of the Milky Way model, but still not too large such that it would introduce errors into the tidal tensor from GMCs coming too close.
     
\subsection{Implementing GMCs into \textsc{nbody6tt}}
The 16 M67 candidates that are expected to survive their GMC encounters have on average 43.6 registered GMC encounters according to Eq. \ref{eq:GMC_reg}. Based on the work done by \citet{Gustafsson2016} we know that GMC encounters that inject less than a few $\delta_E = 0.01$ will not affect the cluster in a significant way. We therefore only focused on the GMC encounters with $\delta_E \ge 0.01$ when it came to implementing them into the $N$-body simulation. The GMC encounters with $\delta_E < 0.01$ were still considered but because of their weaker interactions they are treated as part of the galactic tidal field defined by the tidal tensor. By considering only the stronger GMC encounters we see that our 16 clusters on average have 4.9 GMC encounters. 

\begin{figure}
 \includegraphics[width=\columnwidth]{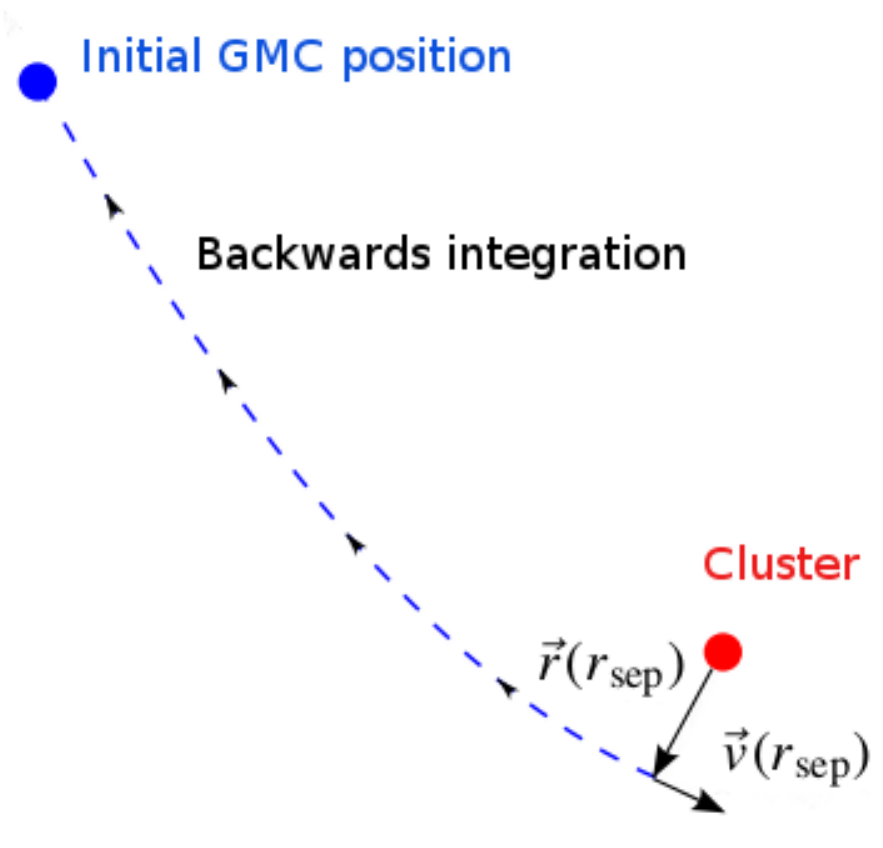}
 \caption{Picture of how the initial position and relative velocity of each GMC was found from the Milky Way model information. By using the minimum separation between the cluster and GMC $r_{\mathrm{sep}}$, the relative position $\vec{r}(r_{\mathrm{sep}})$, the relative velocity $\vec{v}(r_{\mathrm{sep}})$ and time of separation $t(r_{\mathrm{sep}})$ we were able to integrate the relative orbit of the GMC and cluster back to obtain the relative initial position and velocity of the GMC. This was needed since the mass of the GMC changes as a function of time.}
 \label{fig:art}
\end{figure}

The GMCs in \textsc{nbody6tt} behave in a similar fashion as in the Milky Way model in terms of being a \citet{Plummer1911} potential with a mass evolution and core radius according to Eq. \ref{eq:M_t} and \ref{eq:R_c}, respectively. From the Milky Way model we record the minimum separation between the cluster and GMC during each encounter, $r_{\mathrm{sep}}$, the relative position $\vec{r}(r_{\mathrm{sep}})$, the relative velocity $\vec{v}(r_{\mathrm{sep}})$ and time of separation $t(r_{\mathrm{sep}})$. Using this information we integrate the relative orbit of the GMC and cluster back to the birth time of the GMC to obtain the relative initial position and velocity of the GMC. This integration was needed because the mass of the GMC changes as a function of time. A picture of the scenario of the backwards integration can be seen in Figure \ref{fig:art}.   

\subsection{Escapers}
The treatment of stellar escapers in nbody6 is normally done by demanding that the stars have a postive orbital energy and are located more than 2 tidal radii away from the cluster. Since we can have compressive tides in \textsc{nbody6tt}, the tidal radius is no longer a good criterion for our escapers \citep{Renaud2011}. The criteria we instead adopted for a star being an escaper was that it should have a postive orbital energy and be at least ten half-mass radii from the cluster.        

As the $N$-body simulations of the M67 candidates evolve we record the escape time, $t_{\mathrm{esc}}$, and velocity, $\vec{v}_{\mathrm{esc}}$, for each stellar escaper. Each cluster was then re-run in the Milky Way model, where the orbit was integrated with the same initial values and GMCs as earlier. Hence the cluster follows the same orbital path as before. The stars escape the cluster as test particles following their escape history from the $N$-body simulation. As the stars escape they will still feel the gravitational force from the cluster, which is itself evolving in mass and half-mass radius. We represent the force from the cluster as a \citet{Plummer1911} potential with a scale paramter equal to $a = r_h/1.305$, with a linear interpolation of the mass and half-mass radius as a function of time from the results of the $N$-body simulation. 

\section{Results}
\label{sec:Results}
In this section we first focus on a specific model cluster, which we refer to as A, and then later give a more generalized view of all the clusters and trends we see. Cluster A was chosen because it resembles M67 today, and has one of the highest fractions of escapers that end up on solar-like orbits, $f_{\odot}$.

\subsection{Cluster A}
\begin{figure}
 \includegraphics[width=\columnwidth]{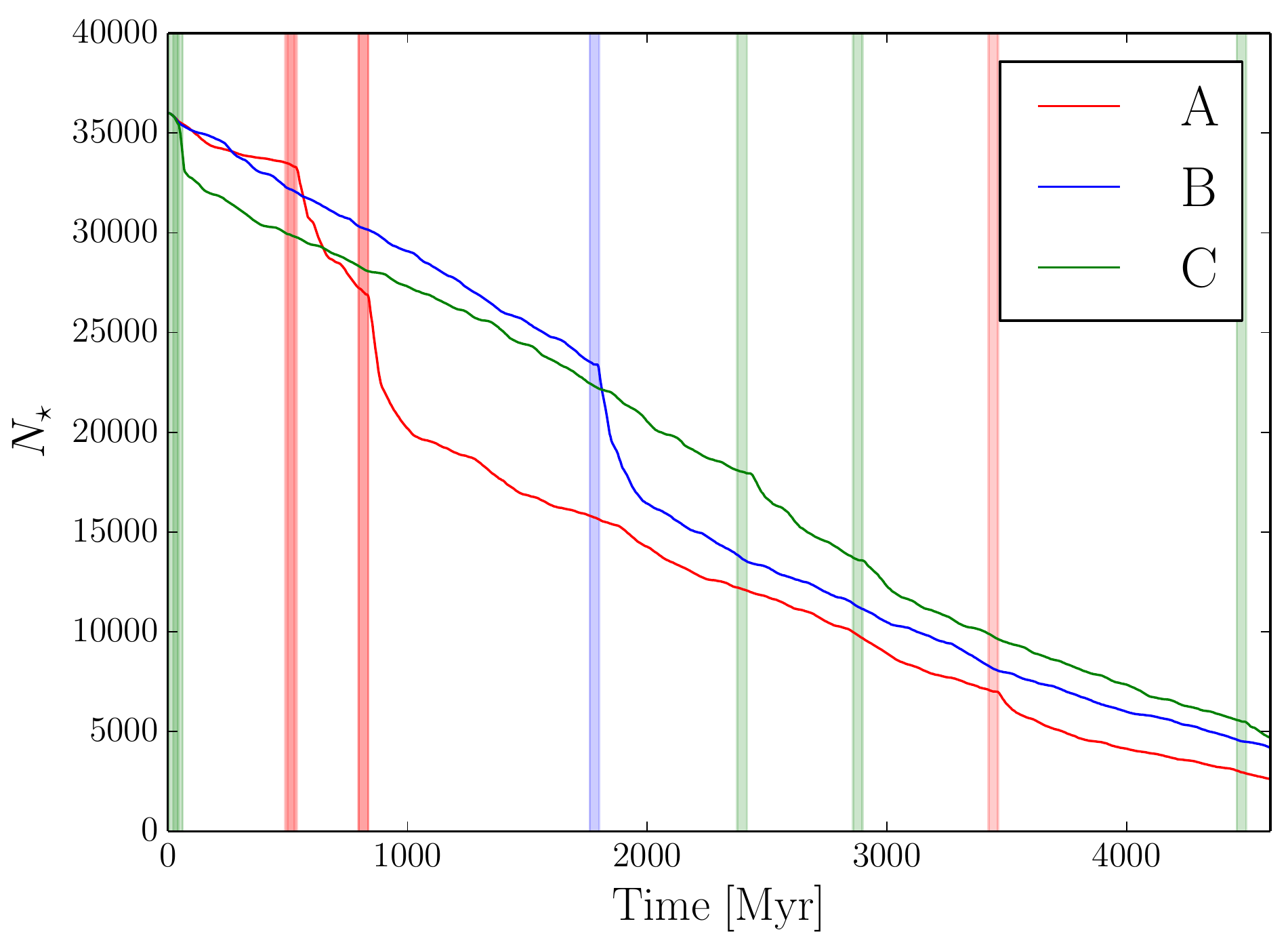}
 \caption{The number of stars for clusters A, B and C as a function of time from the $N$-body simulations. The shaded regions represent the GMC encounters for each cluster; the widths show the lifespan of each GMC. For Cluster A, the first two shaded regions at $t \sim 500$ and $t \sim 800$ Myr encompass two encounters each; hence these shadings appear more pronounced than the final encounter.}
 \label{fig:N_ABC}
\end{figure}

\begin{figure}
 \includegraphics[width=\columnwidth]{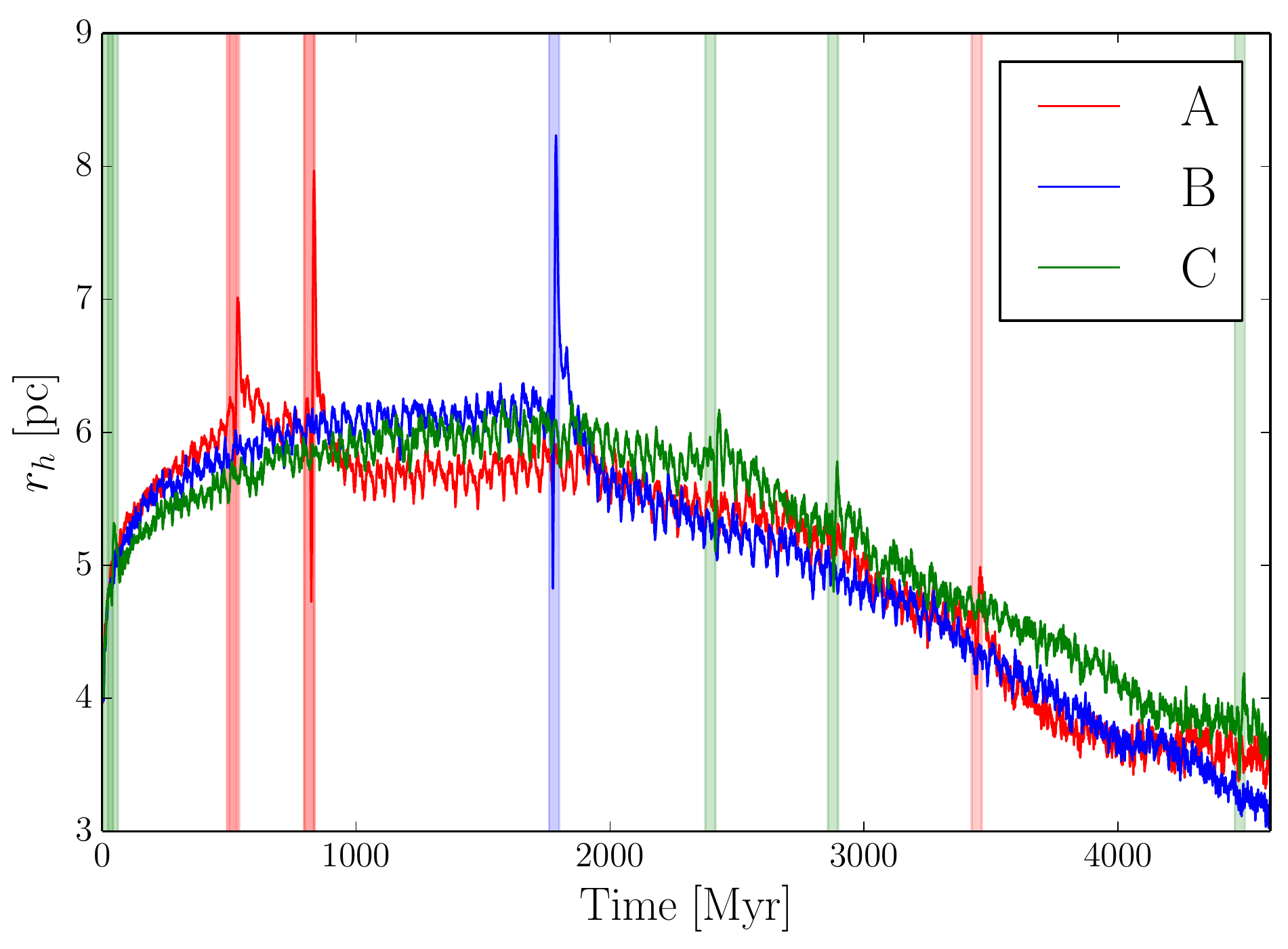}
 \caption{The half-mass radius, $r_h$, as a function of time for clusters A, B and C. Cluster B only has one GMC encounter, but it is also the strongest and causes the largest expansion of the half-mass radius of all three clusters to a value of around 8.2 pc. }
 \label{fig:rh_ABC}
\end{figure}

Cluster A's orbital evolution consists of five GMC encounters, four of which occur before the cluster is 1 Gyr old. The first two encounters are only seperated in time by 2 Myr with encounter times of 518.8 and 520.8 Myr. Similarly the two next encounters also happen in a pair with only 0.9 Myr between them, at times of 819.8 and 820.7 Myr. These four GMC encounters cause the cluster to lose a significant number of stars which can be seen in Figure \ref{fig:N_ABC}. Here the number of stars in the cluster can be seen, together with vertical shaded regions that represent the lifetime of each GMC for each GMC encounter. Roughly 40 \% of the stars are in fact lost within the first 1 Gyr which means that the stellar escapers are leaving the cluster while it is still in a cold disc-like orbit. Between GMC encounters, e.g. in the range of $t = [1000,3300]$ Myr, we see a constant loss of stars that is a consequence of the internal dynamics of the cluster and the external galactic tidal field. In this time between the GMC encounters, the cluster loses around 5 stars per Myr.
As the cluster experiences GMC encounters and the galactic tidal field, we see in Figure \ref{fig:rh_ABC} that after the initial expansion, typical of $N$-body models with mass loss, the half-mass radius is fairly constant at a value around 5 pc, from a couple of hundred Myr and untill the cluster is around 3 Gyrs old. The spikes we see in this time span are produced by the GMC encounters which causes the half-mass radius to increase as a consequence of stars being tidally stripped from the cluster. Ultimately these stars escape and the half-mass radius settles down to a value of $\sim 5.5$ pc. After the last GMC encounter in Figure \ref{fig:N_ABC} and \ref{fig:rh_ABC} we still see a constant loss of stars, however the half-mass radius continues to drops down to $\sim 3.8$ pc before stabalizing.

\begin{figure*}
\centering
\begin{tabular}{|c|c|}
 \includegraphics[width=\columnwidth]{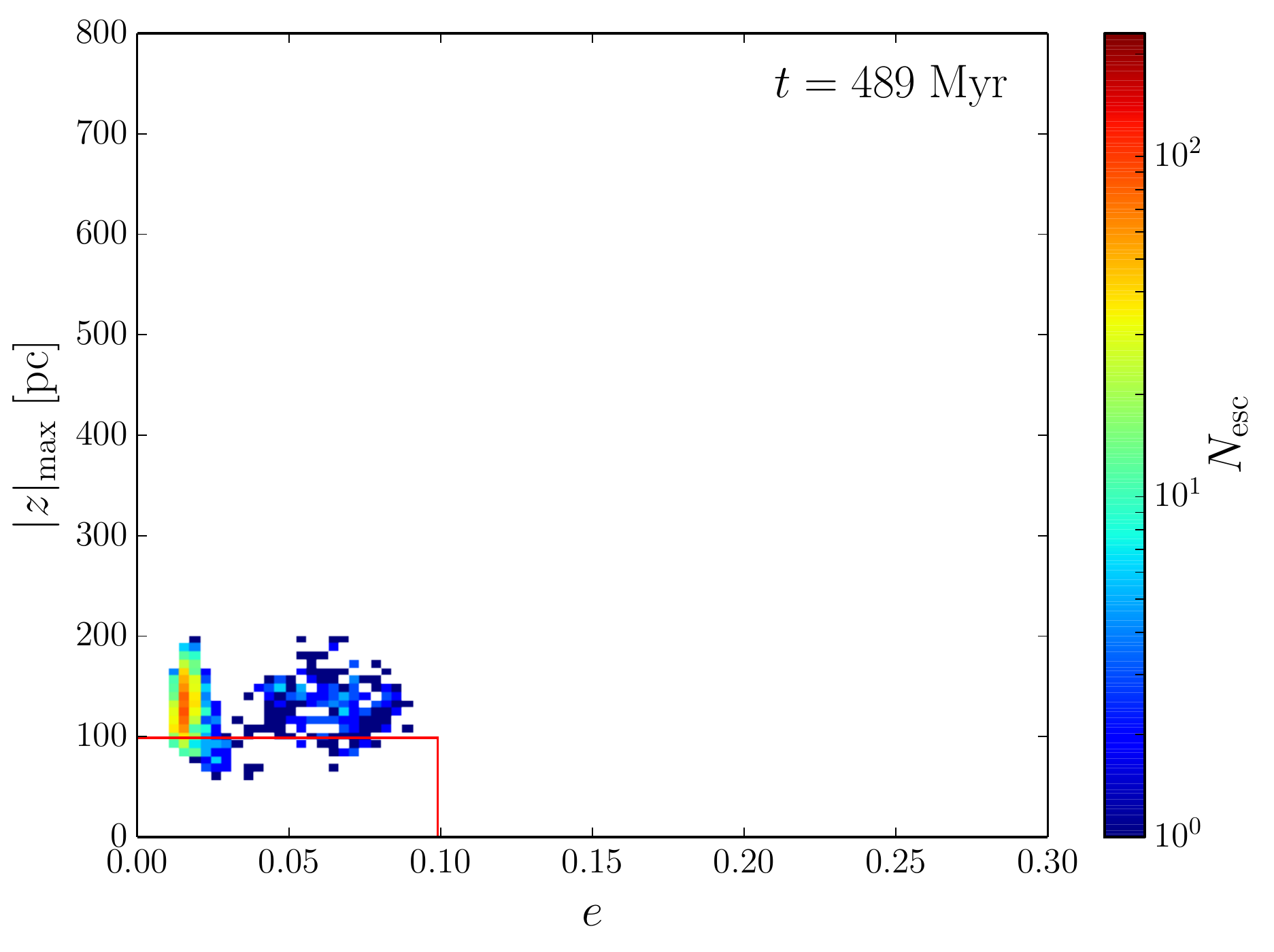} & \includegraphics[width=\columnwidth]{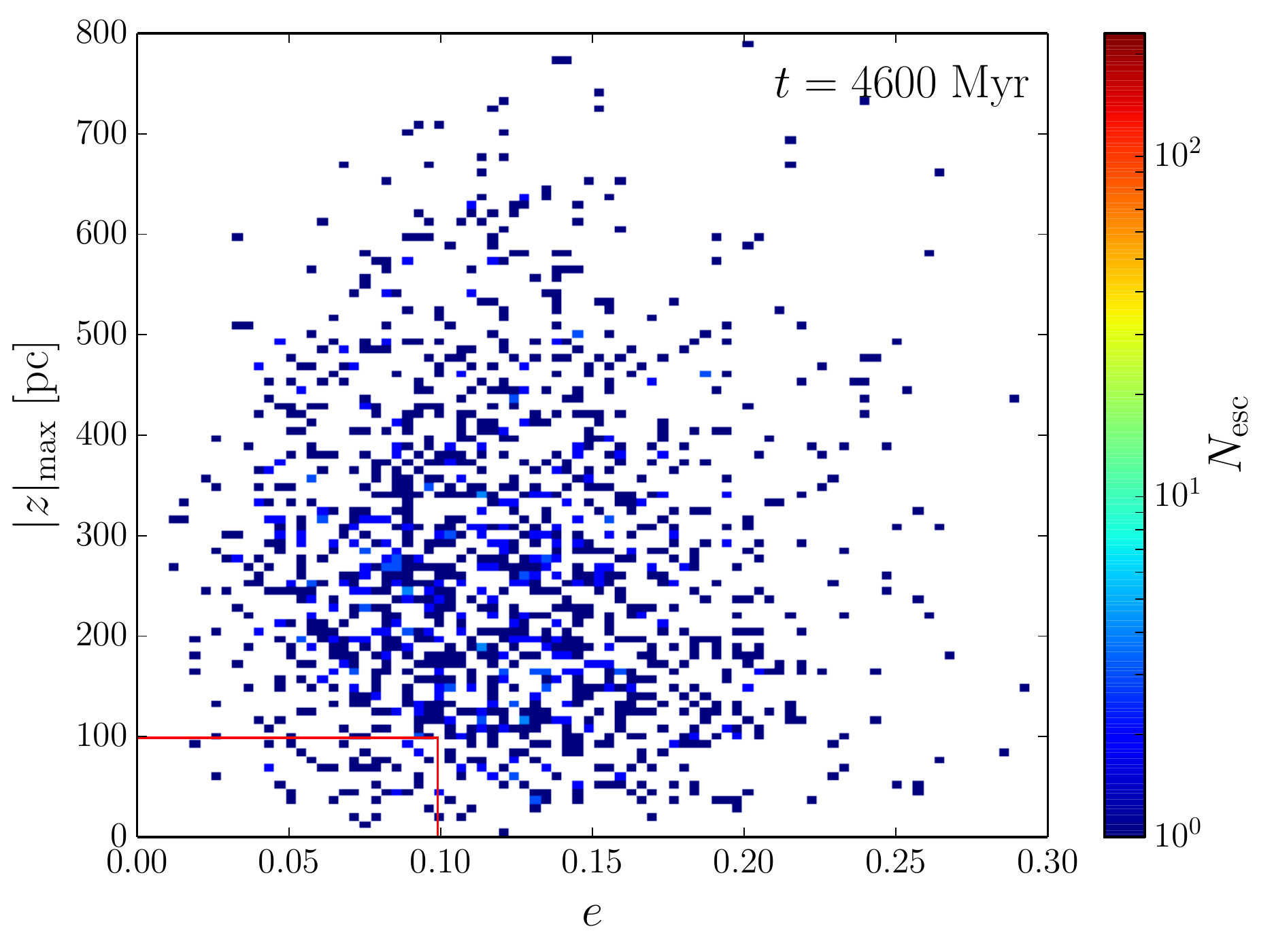} \\
 \includegraphics[width=\columnwidth]{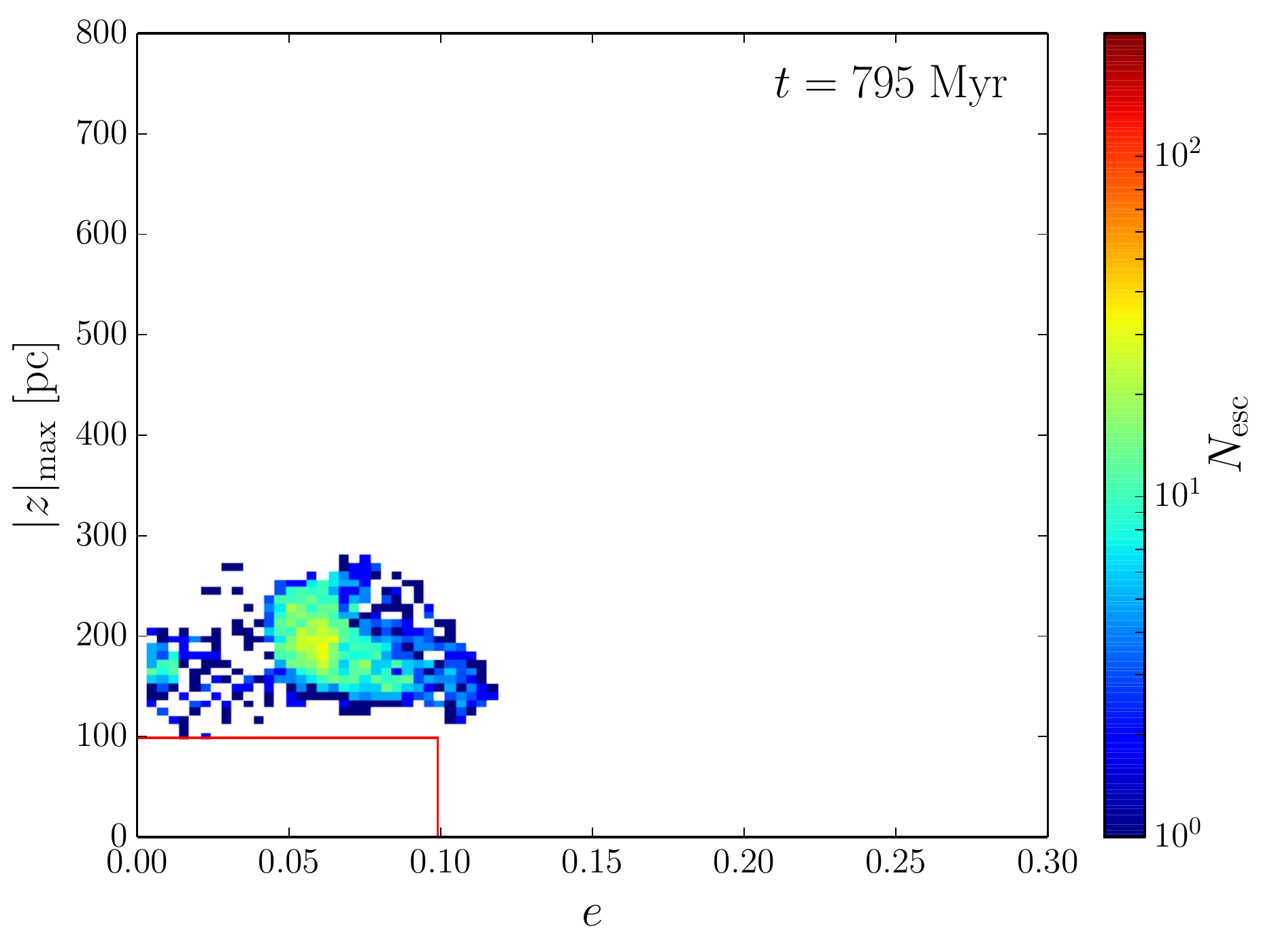} & \includegraphics[width=\columnwidth]{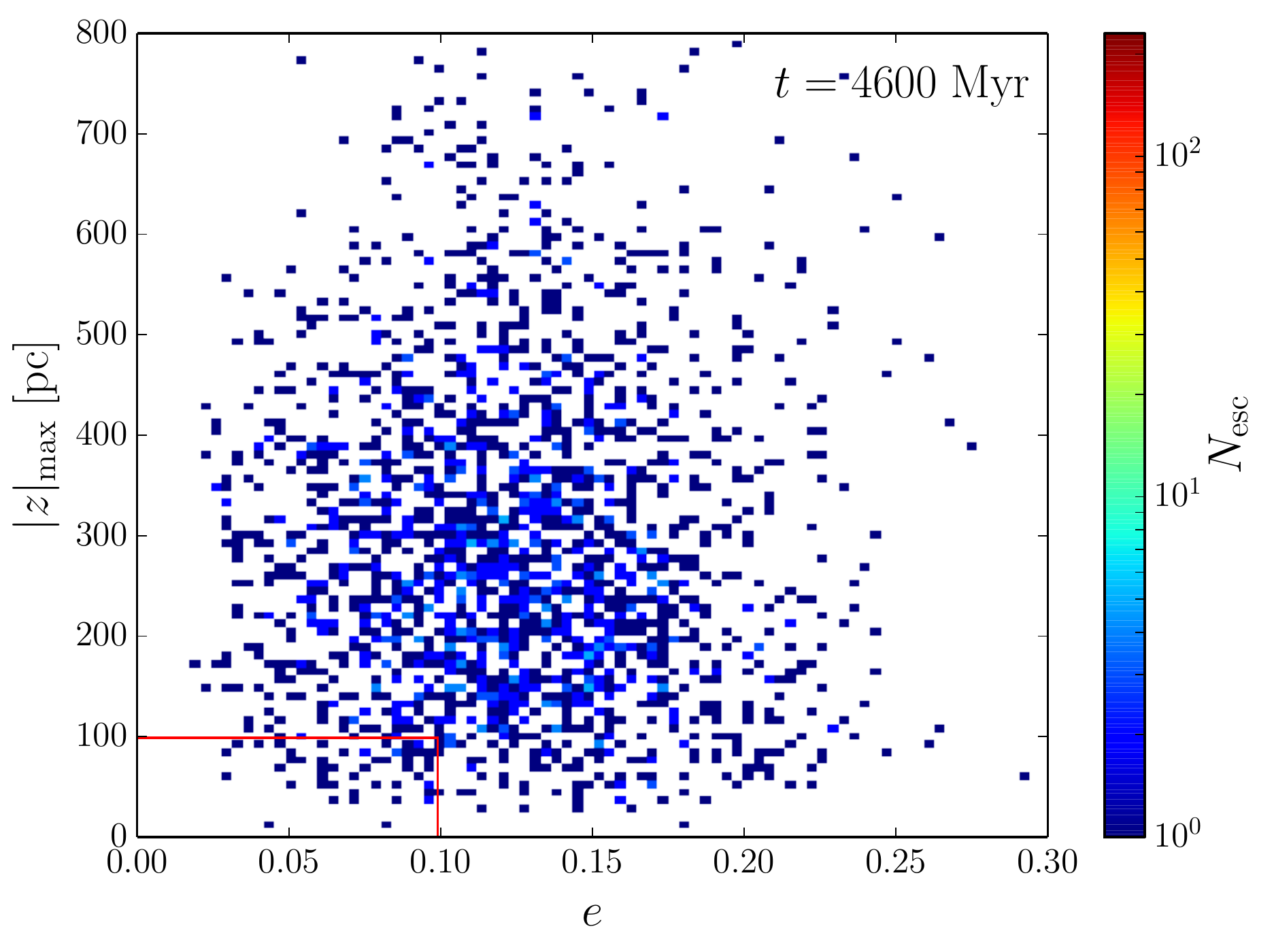} \\	
 \includegraphics[width=\columnwidth]{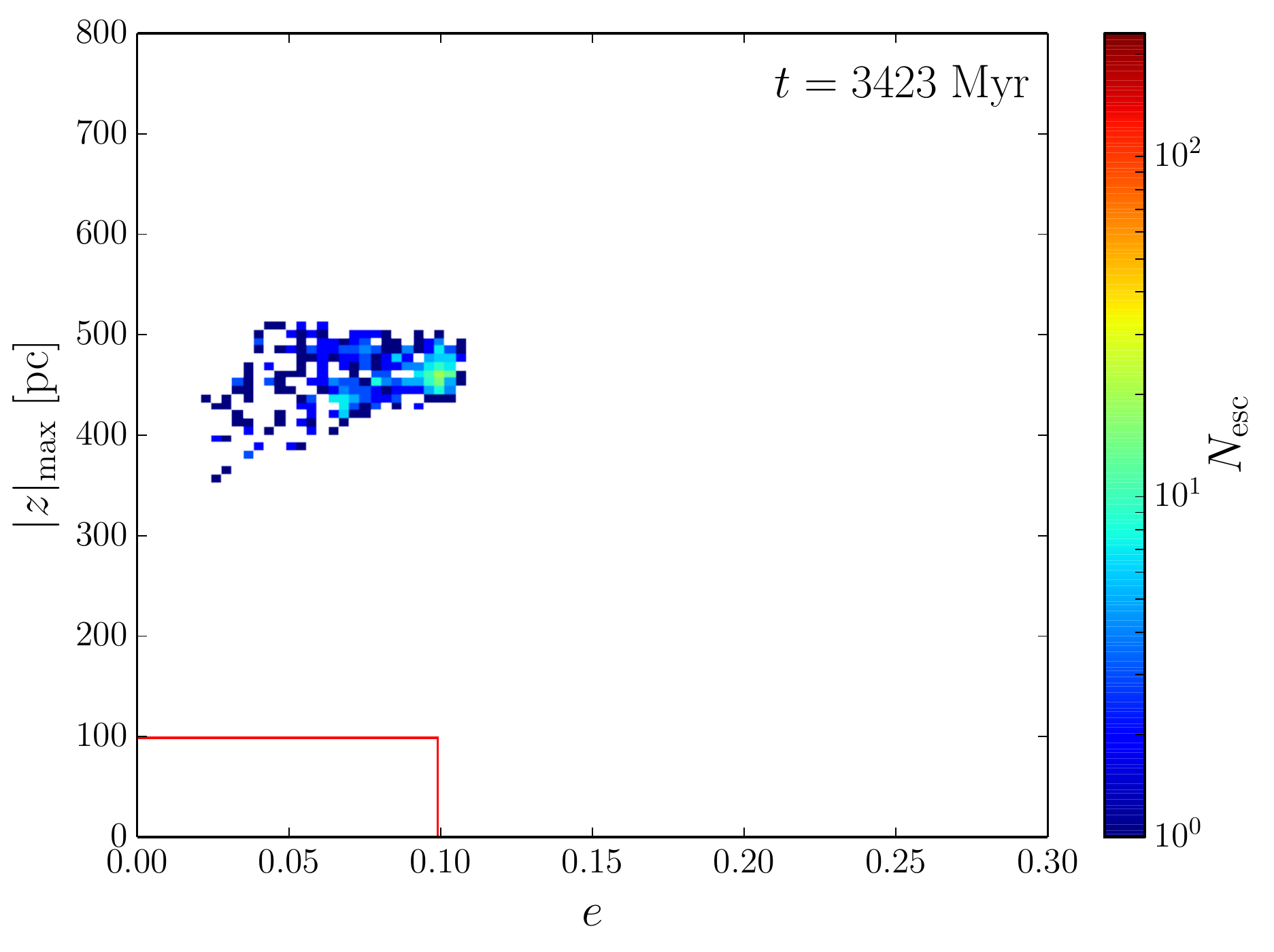} & \includegraphics[width=\columnwidth]{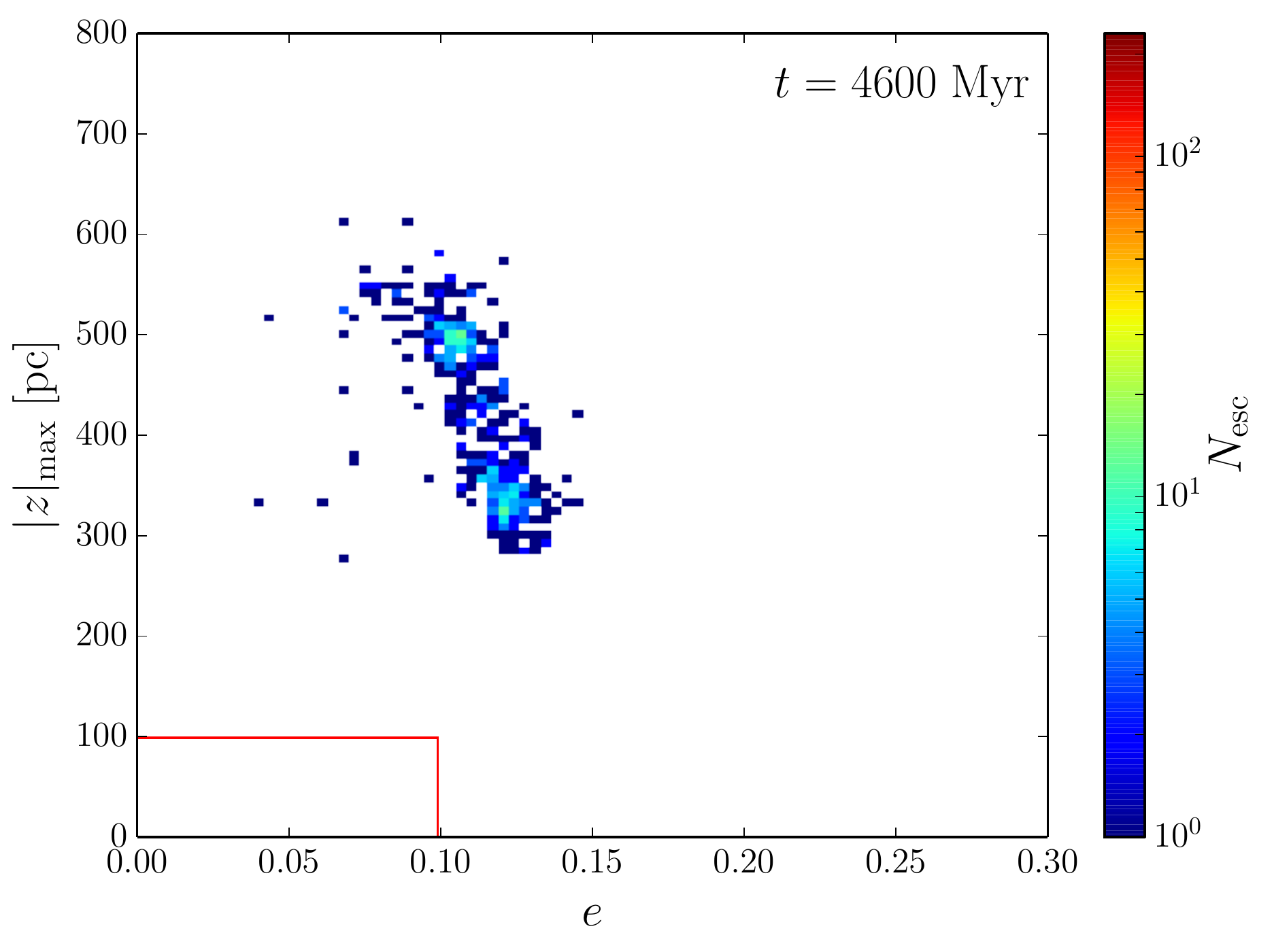} \\
\end{tabular}
\caption{These plots show the orbits of stars that escape cluster A during three separate GMC encounters in the $e$-$|z|_{\rm{max}}$ plane. To the \textit{left} the stars are shown one full orbit around the Milky Way after they escape. To the \textit{right} we see their final orbits at the end of the Milky Way simulation. The red rectangle marks the region within which the orbits defined to be solar-like. From the two first rows of panels, for stars that escape before 1 Gyr, we see that the initial structure of the escapers in $e$-$|z|_{\rm{max}}$ space is completely washed out by the present day due to their interactions with the galactic enviroment.}
\label{fig:PS_A}
\end{figure*}

Because stars escape from cluster A at very different times their orbits evolve separately after they are lost from the cluster. Figure \ref{fig:PS_A} shows the orbits of stars that escape as a consequence of three GMC encounters for cluster A. The \textit{left panels} show the stars at their escape time and \textit{right panels} show the orbits of the same stars at the end of the simulation. From the left panels we see how the initial $|z|_{\mathrm{max}}$ of the escapers depends on when the stars escape. This is a consequence of the orbital heating the cluster undergoes as its orbit evolves in the Galactic enviroment and is scattered to higher $|z|$. The initial orbit of an escaped star is therefore very dependent on what kind of orbit the cluster is on. This is because a majority of the stellar escape speeds relative to the cluster are below 5 \kms \, which we will discuss in later sections. 

From the two first rows of panels, for stars that escape before 1 Gyr, we see that the initial structure of the escapers in $e$-$|z|_{\rm{max}}$ space is completely washed out by the present day due to their interactions with the galactic enviroment. We therefore see no distinction between the final orbits because the stars have been lost early enough in the cluster's history to be equally affected by the galactic enviroment. For the last encounter that happens at $t = 3423$ Myr we see that the escapers have not had enough time to fully diffuse out from the  cluster's original orbit. These stars also escape the cluster at a time where the cluster is already at a high $|z|$ orbit which means that the rate of interactions with spiral arms and GMCs are significantly decreased, causing the diffusion of escapers to be slower.
\begin{figure}
 \includegraphics[width=\columnwidth]{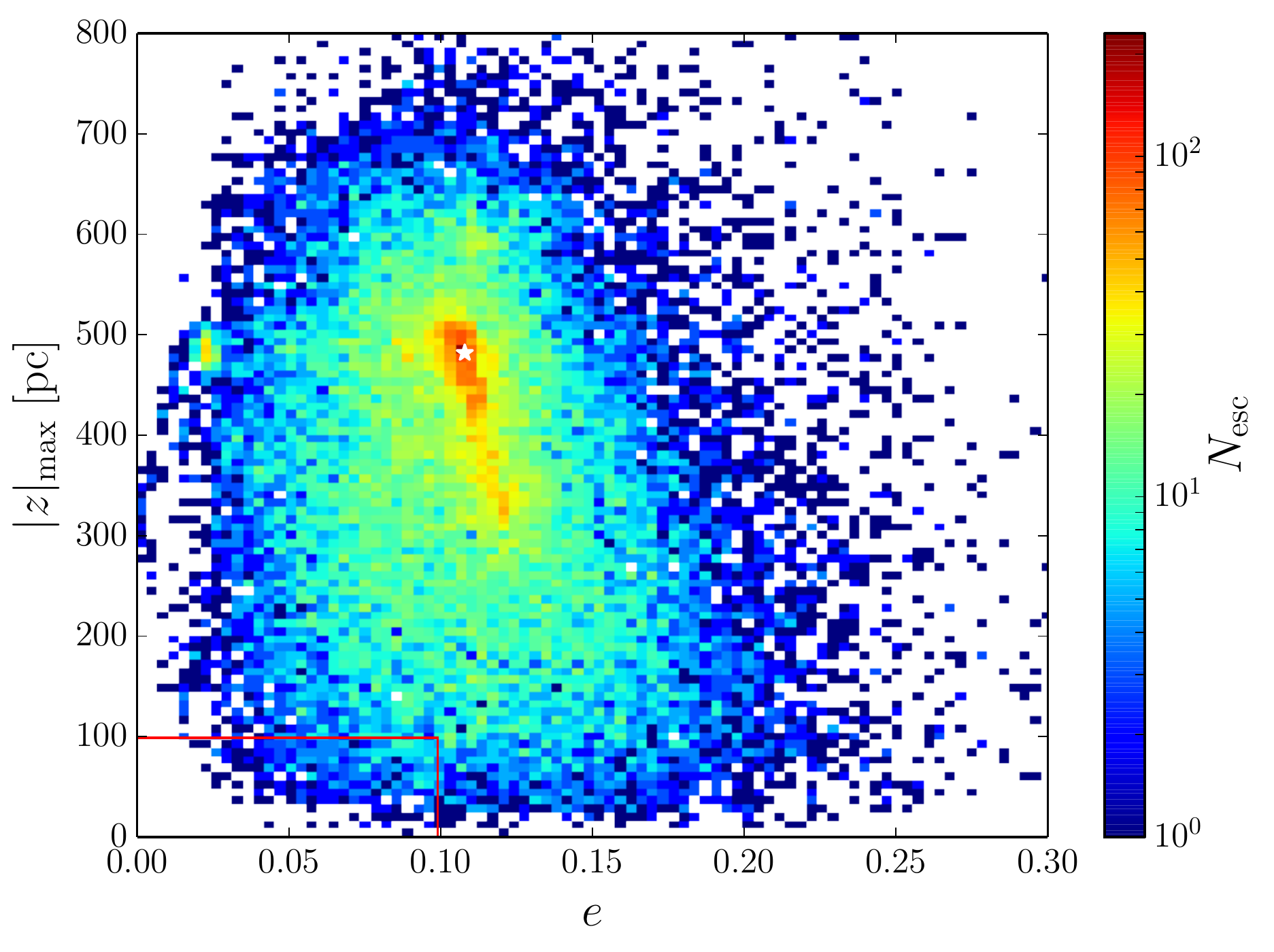}
 \caption{The final orbits of escaped stars from cluster A in the $e$-$|z|_{\rm{max}}$ plane after 4600 Myr. The white star represents the location of cluster A while the red rectangle marks the region where the orbits are defined to be solar-like.}
 \label{fig:2d_hist}
\end{figure} 
Figure \ref{fig:2d_hist} shows the present day orbital parameters of all the escapers from Cluster A. The white star represents the current orbit of M67 and the red rectangle indicates the area for which we defined the stars to be on a solar-like orbit. We see a large concentration of stars around the cluster, which indicates that the stars that have escaped recently, still have orbits similar to the cluster's. This makes sense, since at later times the clusterwill be at high altitude and therefore the interaction with the Milky Way disc will be less frequent and the diffusion of the stars from the cluster will take longer. About 1.67 \% of stars that escape from cluster A end up on solar-like orbits, the second largest fraction for the clusters that still have a reasonable number of stars left. The reason for cluster A's numerous solar-like escapers can be found when investigating when the cluster experiences its GMC encounters.

\begin{figure}
 \includegraphics[width=\columnwidth]{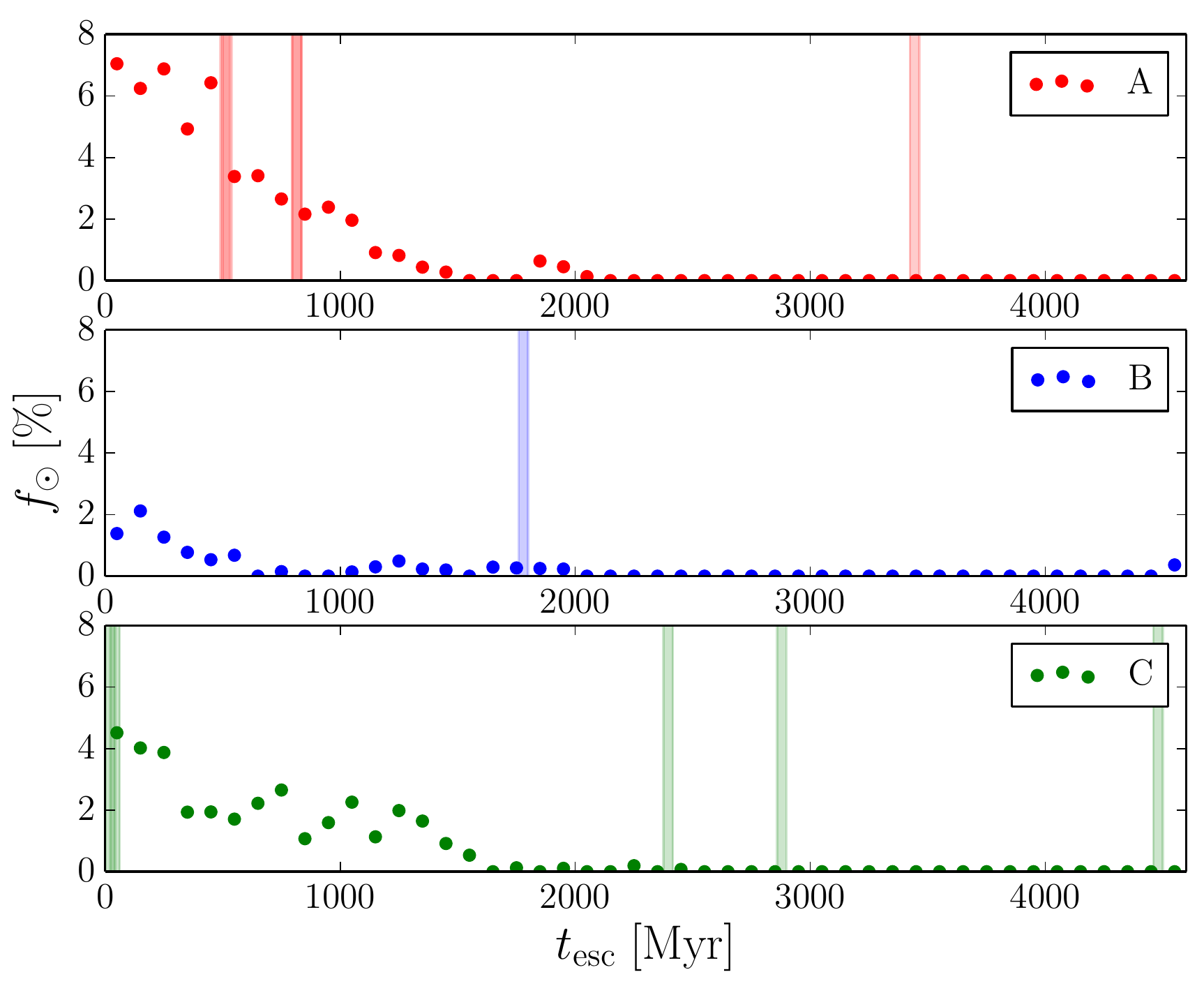}
 \caption{The fraction of stellar escapers that end up on solar-like orbits, $f_{\odot}$, as a function of escape time $t_{\rm{esc}}$. That data are binned in $t_{\rm{esc}}$ with bin size $\Delta t = 100$ Myr. The figure shows clusters A, B and C.}
 \label{fig:frac}
\end{figure}

\begin{figure}
 \includegraphics[width=\columnwidth]{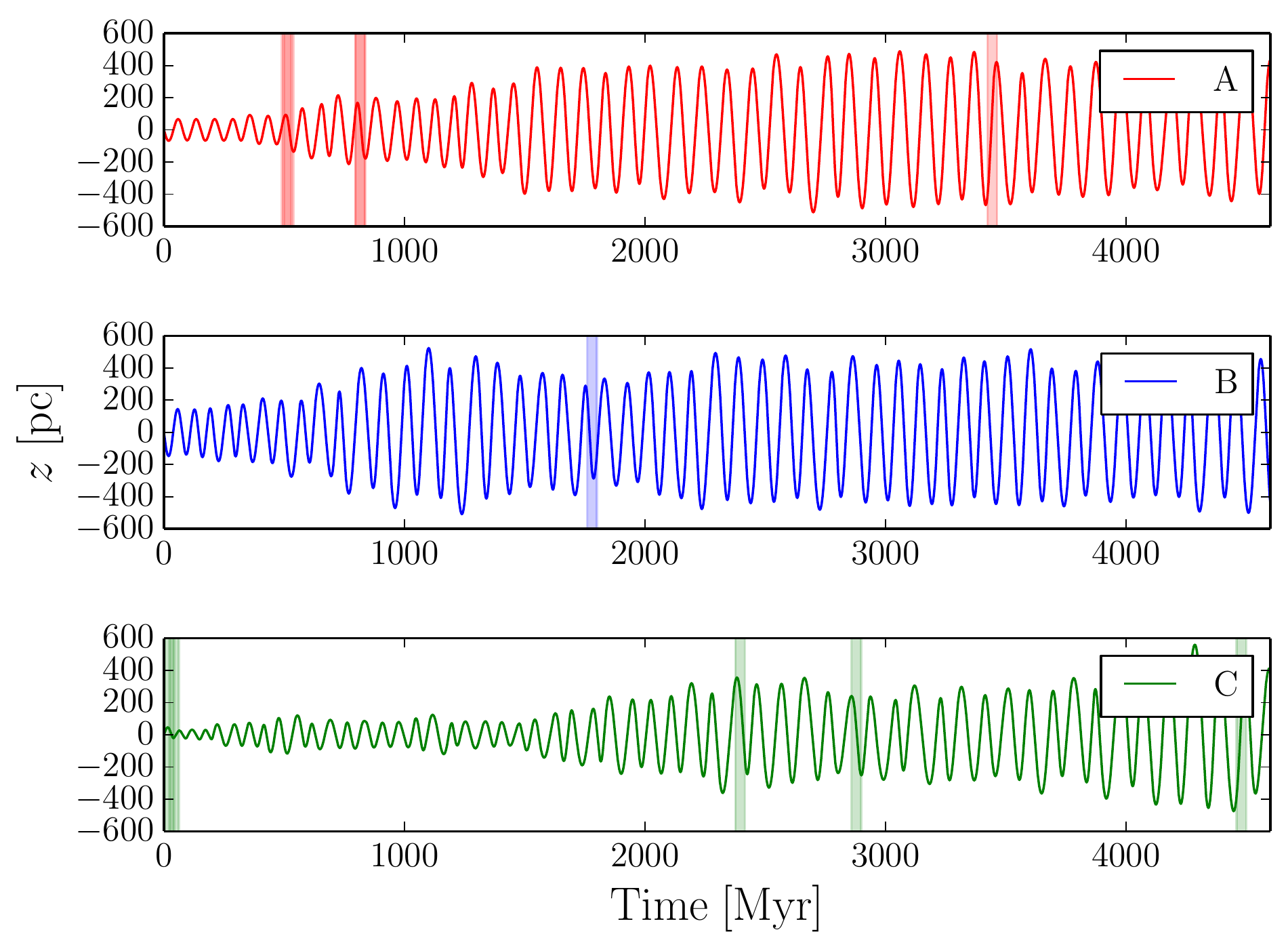}
 \caption{The evolution of the $z$-position as a function of time for clusters A, B and C. Cluster B starts off on a higher $z$-velocity orbit compared to cluster A and C which means that cluster B only experiences one GMC encounter.}
 \label{fig:Xt}
\end{figure}

\begin{figure}
 \includegraphics[width=\columnwidth]{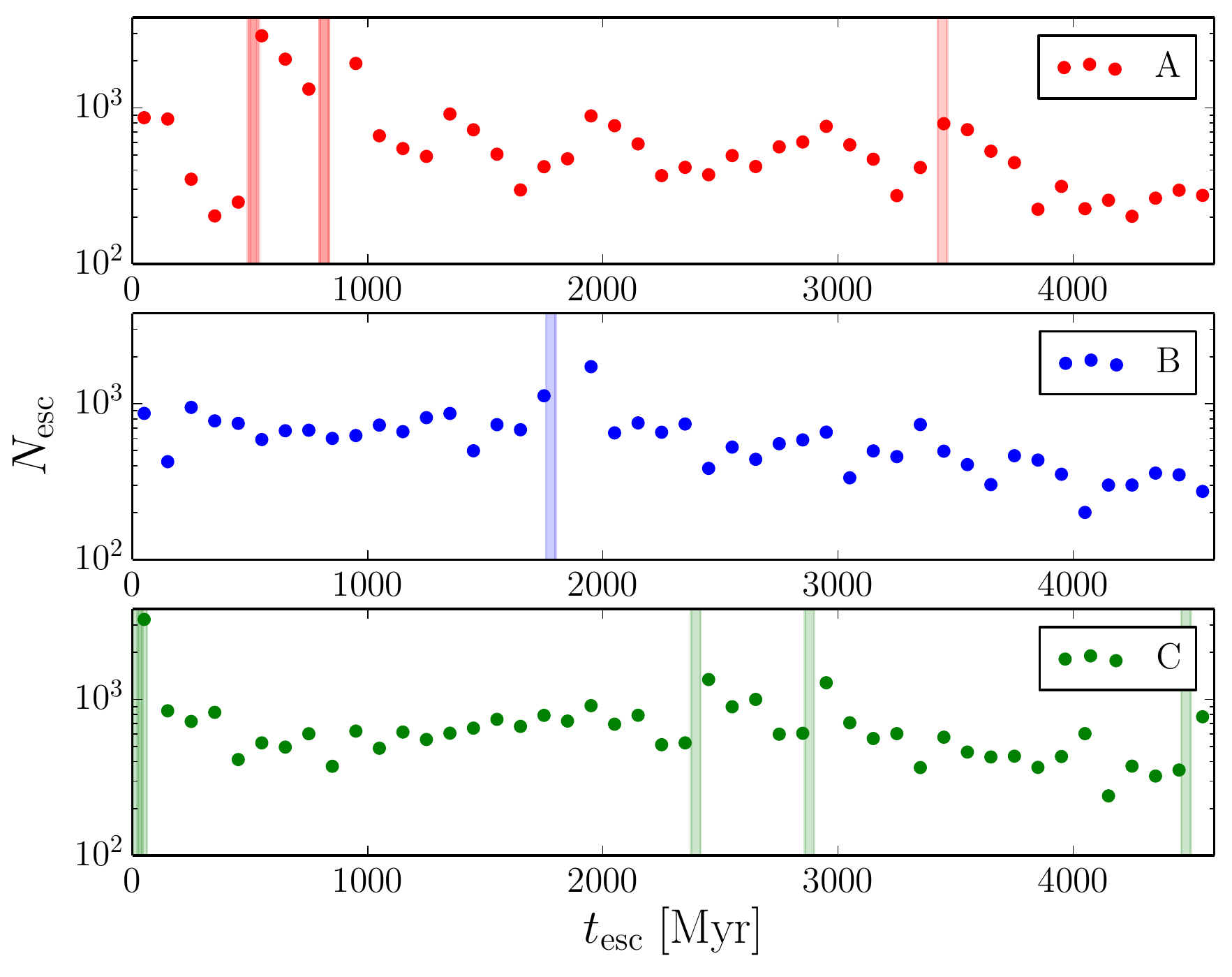}
 \caption{The number of escapers from clusters A, B and C as a function of escape time. The vertical lines indicate GMC encounters which increases the number of escaping stars from the cluster. We also see a periodic trend in the number of escapers with a timescale of $\sim$ 800 - 900 Myr which corresponds well with the timescale of closest contact between the cluster and the spiral arms in the Milky Way model.}
 \label{fig:Nesc}
\end{figure}

Looking at Figure \ref{fig:frac} we see the fraction of escapers that end up on solar-like orbits, $f_{\odot}$. Here $f_{\odot}$ is shown as function of time with a timestep of 100 Myr. We see that $f_{\odot}$ is highest in the begining of the simulation where the cluster is still in a cold disc-like orbit seen in Figure \ref{fig:Xt} which shows the $z$-position of the cluster as a function of time. The GMC encounters do not change this fraction, however they substantially cause a lot of stars to escape during the time while the cluster is still in this cold disc-like orbit. As the cluster is gradually heated by interactions with the spiral arms and GMCs, the fraction of solar-like orbits decreases. After around 2 Gyr the cluster is already at the same height as M67 and we see that the fraction of ejected stars that eventually reach solar-like orbits is non-existent after this point. An important factor for a M67-like cluster to have a high $f_{\odot}$ value seems to be that it has early GMC encounters while the cluster is still in a cold disc-like orbit.    
\begin{table}
 \caption{Results for our 16 M67 candidates. Coulmn list: number of stars remaining in the cluster, half-mass radius, mass, fraction of escapers that end up on solar-like orbits, fraction of escapers that end up on solar-like orbits that have escaped the cluster before the cluster age was 250 Myr, initial $z$-velocity and guiding radius.}
 \label{table:C}
 \begin{tabular}{crrrrrrr}
  \hline
  Cluster & $N_{\star}$ & $r_h$ & $m$ & $f_{\odot}$ & $f_{\odot,250}$ & $v_{z,0}$ & $R_{g,0}$ \\
   &  & $[\mathrm{pc}]$ & $[M_{\odot}]$ & [\%] & [\%] & $[$\kms$]$ & $[\mathrm{kpc}]$ \\
  \hline
  \hline
  A & 2618 & 3.5 & 1707.0 & 1.67 & 0.38 & -5.5 & 8.0\\
  B & 4208 & 3.3 & 2592.0 & 0.25 & 0.09 & -12.3 & 7.4\\
  C & 4683 & 3.6 & 2812.4 & 1.07 & 0.61 & 4.4 & 8.1\\
  D & 183 & 2.8 & 174.7 & 6.61 & 1.03 & -3.6 & 10.0\\
  E & 3209 & 3.3 & 2092.7 & 0.31 & 0.12 & 17.4 & 8.4\\
  F & 3361 & 3.4 & 2172.3 & 0.62 & 0.13 & -8.8 & 7.9\\
  G & 6784 & 3.9 & 3757.4 & 0.49 & 0.22 & 14.7 & 7.4\\
  H & 9837 & 4.6 & 4994.8 & 0.38 & 0.13 & -11.3 & 8.3\\
  I & 774 & 2.8 & 622.1 & 3.74 & 0.47 & -5.8 & 8.0 \\
  J & 5456 & 3.5 & 3184.5 & 0.06 & 0.02 & 20.7 & 8.2\\
  K & 639 & 2.17 & 545.2 & 2.02 & 0.24 & 17.4 & 7.2\\
  L & 1542 & 2.9 & 1141.4 & 0.77 & 0.17 & -2.7 & 7.4\\
  M & 1277 & 2.6 & 980.2 & 5.07 & 0.27 & 1.5 & 9.0\\
  N & 379 & 2.4 & 980.2 & 2.84 & 0.47 & 2.1 & 7.7\\
  O & 256 & 2.3 & 244.4 & 2.10 & 0.64 &  -1.1 & 7.3\\
  P & 2951 & 3.5 & 1933.1 & 1.07 & 0.10 &  -18.7 & 8.4\\  
  \hline
 \end{tabular}
\end{table}

Aside from having a roughly constant stellar escape rate between GMC encounters, Figure \ref{fig:Nesc} shows signs of a periodic signal in the escape rate. Here we see the number of escapers as a function of their escape time which shows a periodic signal with a amplitude of $\sim 250$ stars and a period of $\sim 800 - 900$ Myr. This period agrees well with the timescale of closest approach between the cluster and the spiral arms at the current $R_g$ of the cluster. This signal could therefore be a manifestation of the two spiral arms in the galactic model, however this is hard to verify since the spiral arms are stretched over a large area. From Figure \ref{fig:Nesc} we also see GMC encounters lead directly to the loss of around 10 \% of the cluster's stars per encounter. Even so, the majority of stars are not lost as a direct result of an GMC encounters but instead to the internal dynamics of the cluster and its interaction with the galactic tidal field. Approximately 80 \% of the stellar population is lost this way.

\subsection{Comparison of three cluster histories}
In order to explore the spread of cluster histories in our sample we compare the evolution of three clusters: A, B and C. The three cluster models are chosen based on their high values of $f_{\odot}$, their different GMC encounter histories, and the fact that all three clusters survive. Even though they have quite different evolution in terms of GMC encounters the clusters still end up with roughly the same size and number of stars after 4.6 Gyr. The values of $f_{\odot}$ for the clusters are, however, considerably different which can be attributed to the different timings of these GMC encounters. Results for all 16 M67 candidates can be seen in Table \ref{table:C}.  

Cluster B has one very strong GMC encounter which causes the largest expansion of the half-mass radius of all three clusters to a value of around 8.2 pc seen in Figure \ref{fig:rh_ABC}. Afterwards the half-mass radius steadily decreases until the end of the simulation. When this single GMC encounter occurs the cluster has already been heated up to $|z|_{\mathrm{max}} \sim$ 300 pc as seen in Figure \ref{fig:Xt}. The encounter causes around 8000 stars to escape, but since the cluster is already at a high $|z|_{\mathrm{max}}$ the number of stars that end up on solar-like orbits are considerably lower for cluster B than the two other clusters, with $f_{\odot} = 0.25$ \%. It should be noted that cluster B starts on a hotter orbit than A and C, with $v_{z,0} = -12.3$ \kms \, which corresponds to $|z|_{\mathrm{max}} \sim 200$ pc (see Figure \ref{fig:Xt}). Cluster B has around the same escape rate as A and C, seen in Figure \ref{fig:Nesc}, but because B is born on this hot orbit the fraction of escapers onto solar-like orbits at $t_{\rm{esc}} = 0$ Myr are a factor of 4 and 2 smaller than A and C, respectively. This can be seen in Figure \ref{fig:frac}. 

Cluster C has two GMC encounters within the first 100 Myr of its evolution. This causes the cluster to lose around 3000 stars. At around 2400 Myr the cluster has a third encounter, however, this encounter seems to be particularly weak and we do not see an associated loss of stars escaping. A slight increase in the half-mass radius suggests a small expansion of the cluster due to the encounter. The next encounter happens at around 2900 Myr and like the former encounter it does not seem to have a great effect on the cluster. The last encounter happens just before the simulation ends and thus these escapers end up on similar orbits to the cluster. Just like clusters A and B, cluster C has roughly the same stellar escape rate while it starts out on a cold orbit like A with $v_{z,0} = 4.4$ \kms. The difference between cluster A and C, is that C does not have any early encounters which means that the $f_{\odot}$ in the early stages of the cluster history is lower than A, seen in Figure \ref{fig:frac}. Since this is the most important part of the cluster's evolution in terms of losing stars that end up on solar-like orbits, $f_{\odot}$ is only 1.07 \% for cluster C.     

When looking at Figure \ref{fig:Nesc} we see quite different evolutions. As mentioned before, we see a possible indication of the spiral arms affecting cluster A, however, it is not that clear for cluster B and C. The reason for this might be a result of the difference in the evolution of the eccentricity of the cluster's orbits. Cluster A evolves with a fairly stable eccentricity of 0.08 and thus the spiral arms have a more periodic effect on the cluster. Cluster B has a eccentricity of about 0.14 for its first 1800 Myr until it has its GMC encounter which lowers the eccentricty to $\sim 0.11$ for most of the remaining simulation. Cluster B starts out not showing any periodic signal, but soon after the GMC encounter it starts to show a signal somewhat similar to cluster A even though it appears to be significantly weaker. Cluster C experiences an eccentricity of $\sim 0.15$ for most of its orbit and thus the signals from the spiral arms are scrambled together. The average stellar escaper rates of cluster A, B and C are quite similar when looking at the escape rates between their GMC encounters. The majority of stars lost for all three clusters are a result of the galactic tidal field and the internal evolution which means they end up with relatively similar cluster parameters.

\begin{figure}
 \includegraphics[width=\columnwidth]{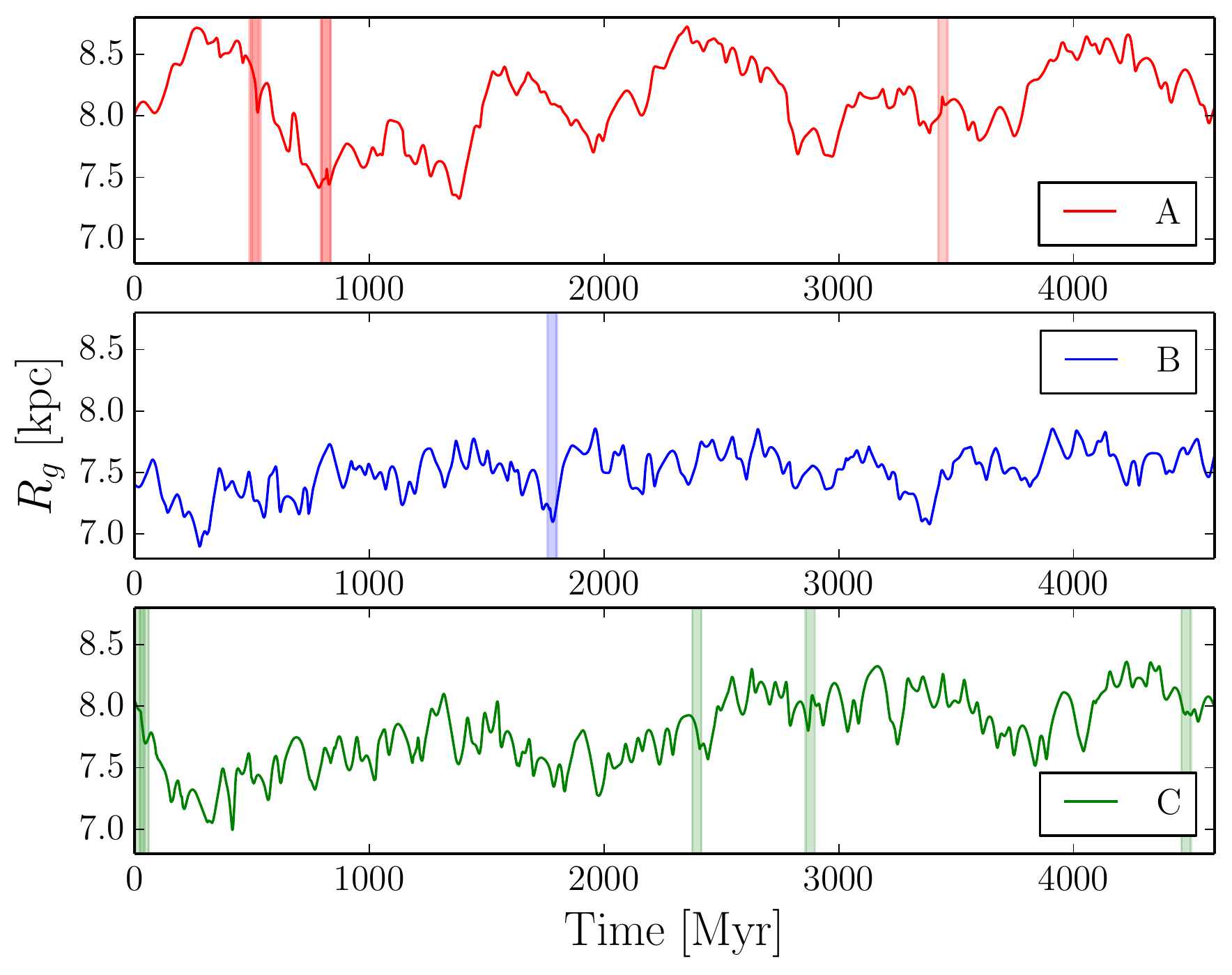}
 \caption{$R_g$ as a function of time for clusters A,B and C.}
 \label{fig:Rg}
\end{figure}

Figure \ref{fig:Rg} shows $R_g$ as a function of time for clusters A, B and C. Cluster A experiences the most churning with changes ranging from 8.7 to 7.4 kpc. Cluster A is therefore more affected by the spiral arms of the Milky Way model which was also our suspicion when looking at the periodic signal in Figure \ref{fig:Nesc}. For clusters A and B we see a small change in $R_g$ at the time of their GMC encounters where as it is not as clear from cluster C. We stated earlier that we did not consider $R_g$ when classifying what a solar-like orbit was in our Milky Way model. Figure \ref{fig:Rg} shows exactly why this would not make sense to do, since the churning of the cluster and stellar escapers depend on what spiral structure is present and if it is transient or not.

\subsection{Trends across all 16 models}
\begin{figure}
 \includegraphics[width=\columnwidth]{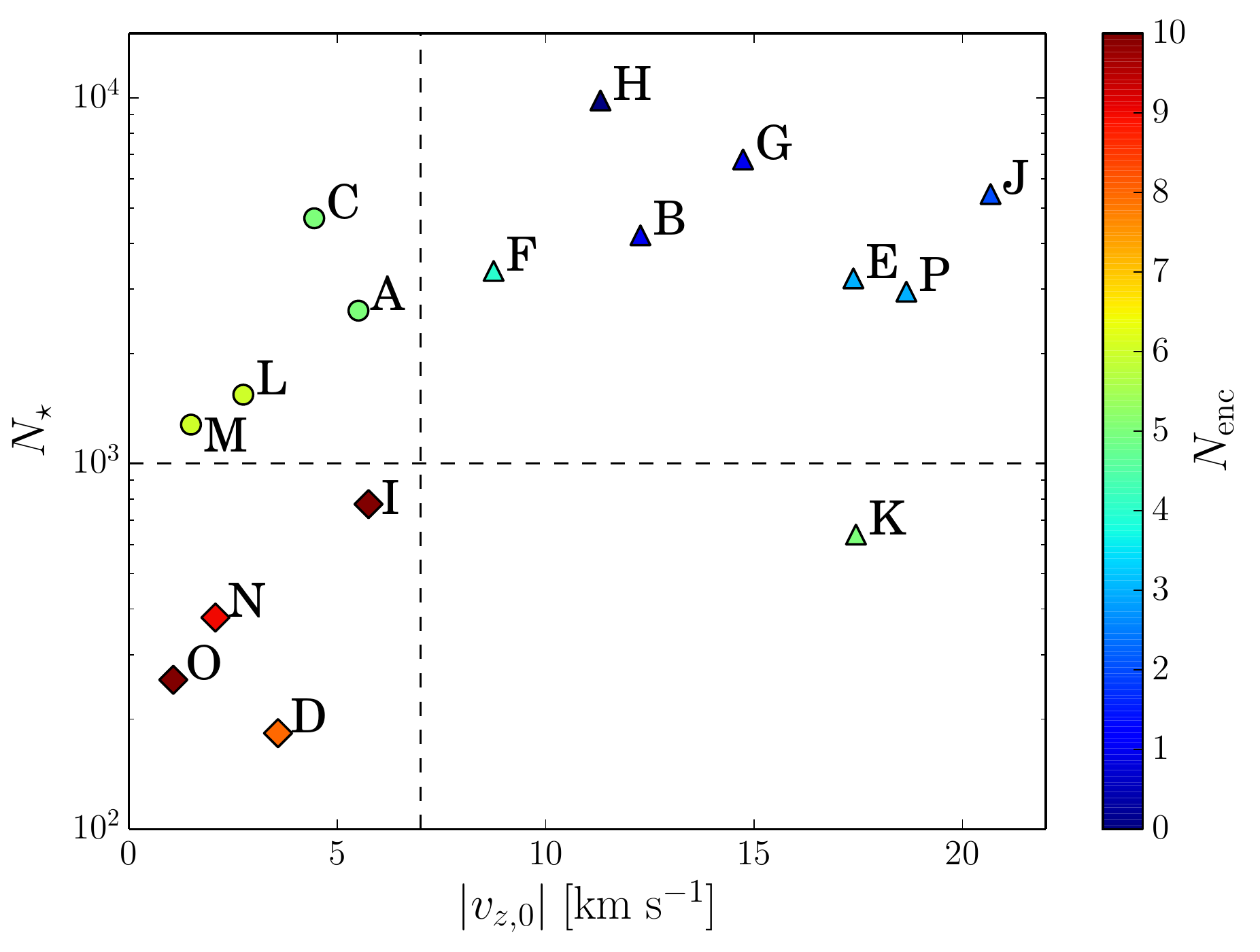}
 \caption{The number of stars in the 16 M67 candidates as a function of their initial $z$-velocity. The vertical line marks the 7 \kms \, corresponding to the velocity dispersion used in the Milky Way model. The vertical line marks the 1000 $N_{\star}$ threshold where we define M67 candidates to be depleted of stars to a point where they do not represent the current number of stars in M67.}
 \label{fig:Nenc}
\end{figure}

We now expand our view and look at all of the 16 M67 candidates. In Figure \ref{fig:Nenc} the number of stars in the cluster after 4600 Myr is shown as a function of the initial $z$-velocity of the cluster. Each cluster is marked and colour coded in regards to the number of GMC encounters it experiences. The vertical line indicates the initial 7 \kms \, dispersion used in the Milky Way model. The vertical line marks the $N_{\star} = 1000$ threshold below which we define M67 candidates to be depleted of stars. The threshold is roughly based on the observed luminous mass of M67 of 1270 $M_{\odot}$ \citep{Fan1996}. From Figure \ref{fig:Nenc} we can divide the 16 M67 candidates up into three groups labeled with the following symbols: 
\begin{itemize}
\item[$\triangle$] \textit{Hot clusters} - Clusters that are born with a high initial $z$-velocity.
\newline
\item[$\Diamond$] \textit{Depleted clusters} - Clusters that are born on cold orbits but are destroyed by GMC encounters in the disc.
\newline
\item[$\circ$] \textit{Scattered clusters} - Clusters that are born on cold orbits with $N_{\star} \ge 1000$ at an age of 4.6 Gyr.
\end{itemize} 
Given that the initial $v_z$ distribution of our cluster particles was Gaussian with $\sigma_z = 7$ \kms, we can see that there is a bias towards clusters born with high $|v_{z,0}|$. Hot clusters get to high $z$ without having to go through as many GMC encounters and risk getting destroyed. Depleted clusters and scattered clusters are in a sense more interesting since they are not outliers in the initial $|v_{z,0}|$ distribution. These clusters all start out with a low $|v_{z,0}|$ and therefore need more heating in terms of GMC encounters to get to the present day $z$-position of M67. We see that the depleted clusters go through the most GMC encounters, between 8 and 10, and as a result they are depleted of stars to the point where they do not match M67. The scattered clusters are born on dynamically cold orbits and manage to survive their 3 - 5 GMC encounters. 

\begin{figure}
 \includegraphics[width=\columnwidth]{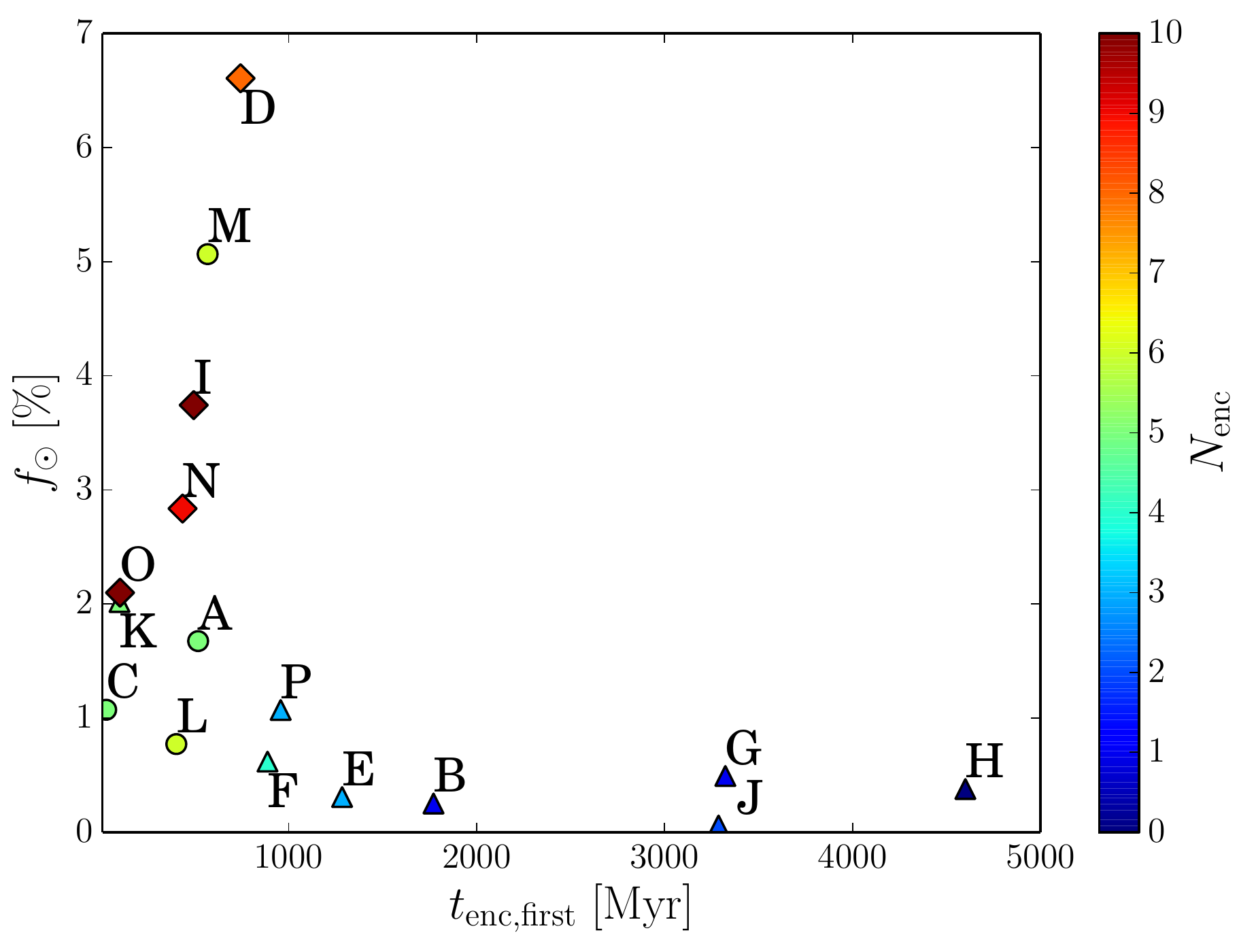}
 \caption{The fraction of stellar escapers that end up on solar-like orbits compared to the time of their first GMC encounter. The clusters are colour coded in regards to the number of their experienced GMC encounters. Hot clusters are marked as triangles and are born with a high $v_{z,0}$, depleted clusters are marked as diamonds and are considered to be destoyed and circles are scattered cluster that represent M67 most in terms of remaining number of stars.}
 \label{fig:enc_first}
\end{figure}

Figure \ref{fig:enc_first} shows the 16 M67 candidates with $f_{\odot}$ as a function of the time of their first GMC encounter, $t_{\mathrm{enc,first}}$. The colour coding and symbols are identical to those in Figure \ref{fig:Nenc}. We see that hot clusters all take longer than 900 Myr before they experience their first GMC encounter and have $f_{\odot}$ below $\sim 1$ \%, except for cluster K which is an outlier with $f_{\odot} = 2.02$ \%. Cluster K experiences an early GMC encounter compared to the other hot clusters and also loses significantly more stars, which could explain the higher value of $f_{\odot}$. Depleted clusters and scattered clusters all experience their first GMC encounter before 1 Gyr and range in $f_{\odot}$ from 0.77 to 6.61 \%. From Figure \ref{fig:Nenc} and \ref{fig:enc_first} there seems to be an anti-correlation between the initial $|v_{z,0}|$ and $f_{\odot}$. We also see that clusters that start with a high $|v_{z,0}|$ generally take longer before they experience their first GMC encounter because they spend most of their time away from the Galactic disc where GMC encounters occur.

\begin{figure*}
 \includegraphics[keepaspectratio=true, scale = 0.51]{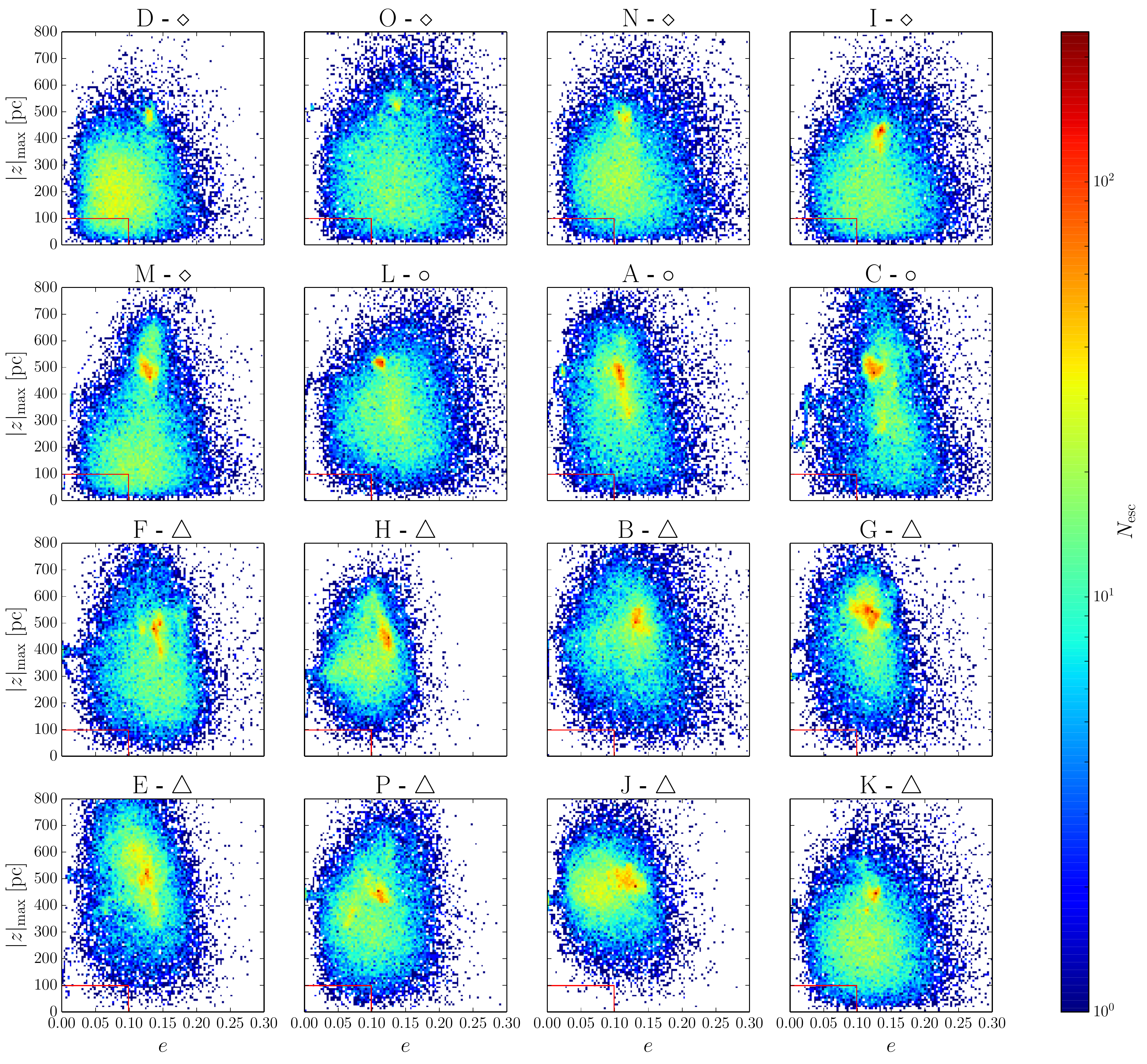}
 \caption{The final orbits in the $e$-$|z|_{\rm{max}}$ plane for the escapers associated with clusters A to P. The clusters are all marked according to their groups and in the order: depleted clusters, scattered clusters and hot clusters.}
 \label{fig:2d16}
\end{figure*}

The final position in the $e$-$|z|_{\rm{max}}$ plane for the escapers associated with clusters A to P can be seen in Figure \ref{fig:2d16}. Depleted clusters are shown first, followed by scattered clusters and hot clusters. For the depleted clusters we see that the escaped stars are located closer to the Galactic disc where the stars are lost due to GMC encounters. For clusters D, O and N we see a lower concentration of escapers around the present day position of M67 indicating that they have lost a higher fraction of their stars before they settled on this orbit. The scattered clusters have their escapers less concentrated to the Galactic disc and a higher concentration around the present day position of M67. This is because they retain more of their stars on their journey towards the present day orbit of M67. The hot clusters have their concentration of escapers higher above the Galactic plane which result in them having a lower $f_{\odot}$ value. They have fewer GMC encounters and less interaction with the Galactic disc which means that the diffusion of stellar escapers is less pronounced for some of the cluster, namely H, G, E and J. Again cluster K, which experiences its first GMC encounter at $\sim 100$ Myr, is an outlier.

\section{Discussion}
\label{sec:Discussion}
In order to compare our Milky Way model to observations and other simulations we can look at Table \ref{table:dispersion}. When comparing our results to the results of \citet{Holmberg2009} we see that $\sigma_U$ and $\sigma_V$ match the observations quite well, while $\sigma_W$ shows the largest discrepancy. As mentioned earlier this suggest that our Milky Way model is biased against test particles going onto high altitudes. For the purpose of our investigation, this means that scattering test particles into M67-orbits in our Milky Way model will be more difficult and thus will most likely underestimate the number of M67 candidates in our model. This does, however, also mean that stellar escapers will experience less vertical scatter which could increase the number of solar-like escapers we see. It should be noted that the results from \citet{Holmberg2009} which used the Hipparcos parallaxes to improve the accuracy of the Geneva-Copenhagen Survey \citep{Nordstrom2004} is not complete to within a distance of 300 pc which we use to classify the Solar-neighbourhood. As stated in \citet{Nordstrom2004} the brightest F stars are complete to a maximum distance of $\sim 70$ pc and G5 stars to within $\sim 40$ pc.  

In comparison, the fraction of test particles that are located at $|z|$ > 400 pc, $f_{400}$, at the end of the simulation is 3.6 \% compared to 1.8 \% from \citet{Gustafsson2016} which can be seen in Table \ref{table:dispersion}. Both of these values have been calculated after taking the initial Galactocentric radius into account and weighing the different test particles following the exponential gas disc of the Milky Way model of $\sim \mathrm{exp}(R_i/4.8 \, \mathrm{kpc})$ as the test particles where uniformly distributed in both simulations. These two values do not agree that well, however, in comparison the simulation of \citet{Gustafsson2016} suffer from small number statistics where 500 and 1000 test particles were used, whereas our simulations are based on five independent runs with a total of 25000 particles. The initial birth radii of the test particles will also have an effect on the discrepancy between our model and the model of \citet{Gustafsson2016}. Their initial birth radii were defined between 4 and 9 kpc whereas ours are between 4 and 10 kpc resulting in a higher fraction of stars being able to reach higher $|z|$ due to the lower strengh of the Milky Way potential at this Galactocentric radius. 

We also simulated a different galactic model (which we will refer to as ROSO from now on) with a different GMC distribution from \citet{Rosolowsky2005} with $\propto M_i^{-1.5}$ making it less top heavy in the mass interval of $4.0 \le \mt{log}(M_i/M_{\odot}) \le 6.5$ resulting in a constant number of 8770 GMCs. All of the velocity dispersions are lower than the values from our standard the Milky Way model and the results from \citet{Holmberg2009}. In particular $\sigma_V$ is lower since the lower mass GMCs are not able to increase the velocity dispersion to match observations. The lower average GMC mass results in a lower $f_{400}$ of 2.1 \% compared to our Milky Way model.
\begin{table}
 \begin{tabular}{ccccc}
  \hline
    & $\sigma_U$ & $\sigma_V$ & $\sigma_W$ & $f_{400}$ \\
    & [\kms] & [\kms] & [\kms] & [\%] \\	
  \hline
  MW model & 31.9 & 18.0 & 11.7 & 3.7\\
  ROSO & 29.9 & 15.7 & 10.5  & 2.1\\
  \citet{Gustafsson2016} & 37.5 & 18.9 & 16.9$^{\spadesuit}$ & 1.8 \\
  \citet{Holmberg2009} & 30.9 & 21.7 & 15.2 & - \\  
  \hline
 \end{tabular}
 \caption{The velocity dispersion of stars in the three galactic rectangular coordinates $U,V$ and $W$ at a time of 4.6 Gyr. A comparison of our Milky Way model with the simulations of \citet{Gustafsson2016} and observations of \citet{Holmberg2009} can be seen. The ROSO simulation are similar to the Milky Way model, but where we use the GMC distribution from \citet{Rosolowsky2005} which is more bottom heavy with a lower average GMC mass. $\spadesuit$: This value was only calculated for stars crossing the Galactic plane.}  	
\label{table:dispersion}
\end{table}

\begin{figure}
 \includegraphics[width=\columnwidth]{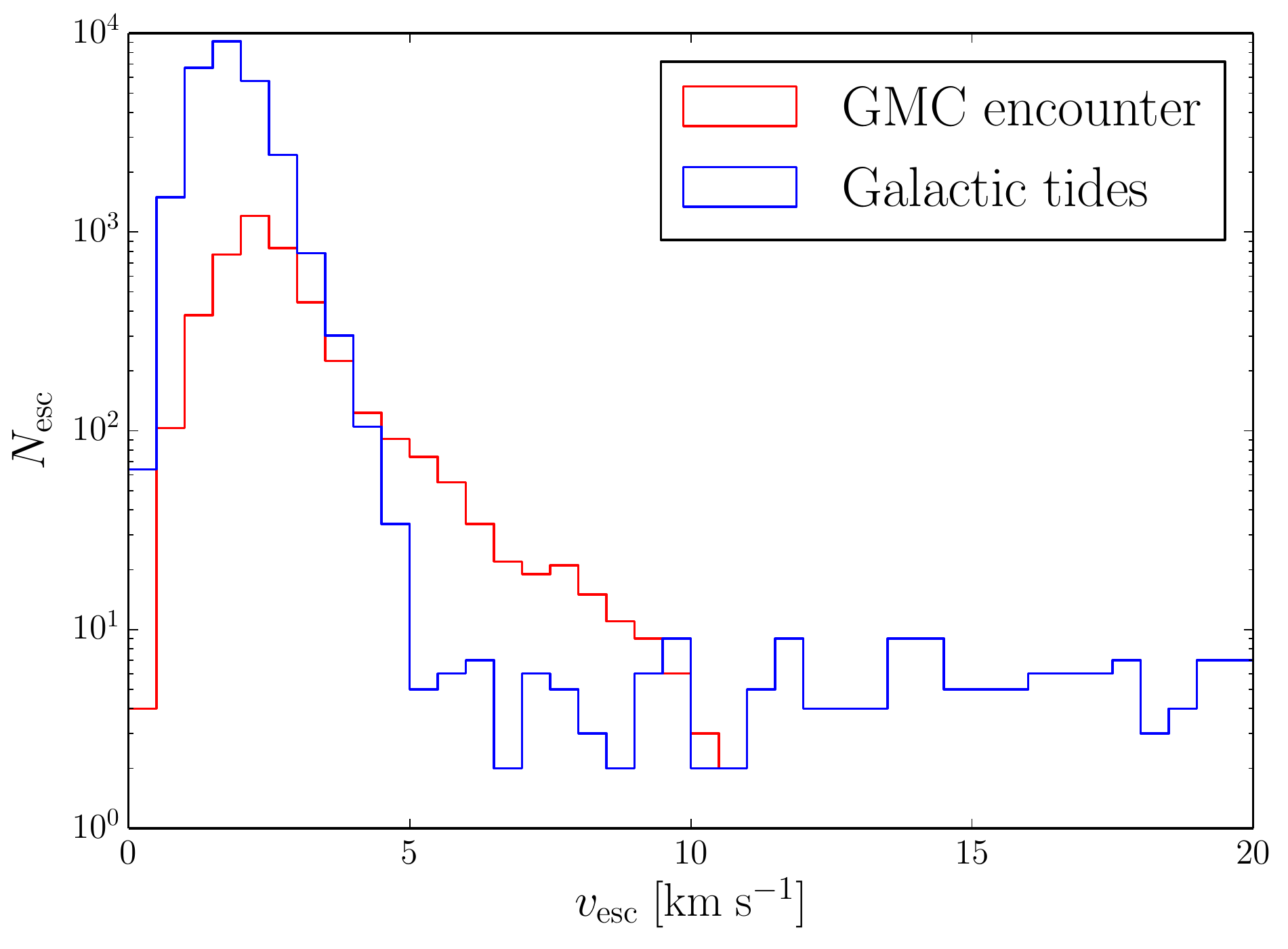}
 \caption{The distribution of ejection velocities for escapers from cluster A. The red histogram shows the escape speed for stars that escape as a direct consequence of an GMC encounter and the blue histogram represents the stars that escape in between these events. For both distributions we see that most of the stars escape with speeds less than 10 \kms, with the majority being below 5 \kms. Stars with ejection velocites above $\sim 10$ \kms \, are all black holes and neutron stars that have received kicks.}
 \label{fig:v_esc}
\end{figure}

\citet{Pichardo2012} have claimed, based on simulations done in a galactic potential very similar to our Milky Way model (but without GMCs), that the Sun could not have been born M67. The arguement presented by \citet{Pichardo2012} was that in order for the Sun to be kicked out of M67 to a very different solar orbit would damage the outer parts of the Solar system. In their simulations they integrated M67 backwards in time from its current location. The result was that the Sun would have to leave M67 with a velocity larger than 20 \kms \, in order for the Sun to end up on the current solar orbit, since this is roughly the $v_z$ by which M67 currently crosses the Galactic plane. Kicks produced by three-body encounters of this magnitude would destroy the outer Solar System and the probability of an GMC being responsible for ejecting the Sun from M67 was calculated to be smaller than $10^{-4}-10^{-5}$. The vast majority of stellar escapers in our simulations all escape with speeds lower than 5 \kms \, which can be seen in Figure \ref{fig:v_esc} for cluster A. Here we have divided the escapers up into two groups representing stars that escape as a consequence of an GMC encounter and stars that escape between GMC encounters due to internal and tidal effects. We see that most of the stars are below 10 \kms \, with the majority being around a few \kms \, for both distributions. We see that GMC encounters cause a wider distribution, but drops off rapidly at $\sim$ 10 \kms \, unlike the stars that escape between GMC encounters. These low escape velocities decreases the risk of destruction of the Solar System compared to the results found by \citet{Pichardo2012} which hardly had any escape speeds below 20 \kms \, which would be enough to cause this destruction. This is because we assume that the cluster is born in the Galactic disc and via interactions with the galactic field and GMCs M67 is scattered to its current orbit and not initially born in it while our escaper orbits also diffuse out from the orbit of the cluster. The relative velocity of around 20 \kms \, is therefore not needed in our scenario. 

To investigate the effect of GMC encounters on the Solar System we examine the tidal work from GMC encounters by using Eq. \ref{eq:d_E} to calculate the fractional change in energy of the orbit of Neptune during a GMC encounter. We pick the most violent GMC encounter occuring for the 16 M67 candidates with $M_t = 4.3 \times 10^6 M_{\odot}$, $r_{\rm{sep}} = 25.8$ pc and $v(r_{\rm{sep}}) = 43.7$ \kms, and adopt Neptune's apocenter of 30.33 AU to maximize the potential destruction of the orbit. We find $\delta_E =8 \times 10^{-8}$ and can therefore conclude that GMC encounters can shape the internal and orbital evolution of stellar clusters, but they will have an negligible effect on the Solar System.   

There is also a time limit when it comes to how long the Sun should be able to stay in a cluster like M67. With a stellar density of $\sim 1000$ pc$^{-3}$, which is significantly higher than our representation of M67, \citet{Adams2010} showed that the Sun would have to leave M67 before the cluster has reached an age of 250 Myr in order not to perturb the orbits of Uranus and Neptune by stellar flybys. Even if we might have a larger time limit, we can use the time limit of 250 Myr as a tight constraint and look at how $f_{\odot}$ changes if we only consider stars that escape the M67 candidates before a time of 250 Myr. This is represented by the parameter $f_{\odot,250}$ in Table \ref{table:C}. We see that clusters such as A, B, C, E and G all lose between 20 to 45 \% of their stellar escapers that end up on solar-like orbits before the time limit of 250 Myr. The majority of our solar-like escapers, are lost early where the cluster is on a colder, disc-like orbit. Even restricting ourselves to stars escaping before 250 Myr we find escapers from our model clusters that reach solar-like orbits at the present day.

The number of GMC encounters for each cluster ranges from 10 down to zero encounters. Some of the clusters only have 0 - 2 encounters and even so, they still reach the height of M67. This is because they are born on dynamically hot orbits with initial $|v|_{z,0} > 7$ \kms. It is not clear whether or not it is in fact possible for a rich M67-like cluster to be born a high $|z|$ orbit, i.e. at the tail of the $z$-velocity distribution. We also see that clusters that are born on dynamically hot orbits have smaller $f_{\odot}$ and therefore make them less likely to be the birth place of the Sun.
We focus therefore on the two other groups of M67 candidates, the scattered clusters and depleted clusters. These are the ones which experience GMCs encounters and survive, i.e. they have a healthy number of stars at the end of the simulation, and the ones which have a low number of stars, effectively meaning that these clusters are destroyed in the disc before they reach a M67-like orbit. It should be noted that the distinction between depleted clusters and scattered clusters is approximate, since there is a continous variation of properties between the clusters in these two groups. From looking at Figure \ref{fig:Nenc} we classify cluster A, C, L and M as scattered clusters and D, I, N and O as depleted clusters. Using this classification we calculate that it is $\sim$ 2 times more likely that a solar-like escaper will be produced by a cluster that gets destroyed in the Galactic disc before it gets scattered up to an M67-like orbit versus a cluster that survives its journey. 

From Figure \ref{fig:enc_first} we see that there is a clear indication that the earlier the cluster has its first GMC encounter, the higher $f_{\odot}$ it ends up with in the end. This is reasonable since the cluster will have a cold orbit in the beginning of its lifetime. Stars that escape early will therefore have a higher probability of ending up on a cold solar-like orbit. Additionally, clusters initially on dynamically hot orbits spend less time in the disc and hence have later first encounters.

The diffusion of the orbital elements of stars that escape cluster A during direct GMC encounters can be seen in Figure \ref{fig:PS_A}. This tells us that the nature of the diffusion is controlled by the galactic gravitational field and that it is only possible to link stars to a specific GMC encounter if the encounter has happened within a few galactic orbits. Stars that escape clusters that are located on high $|z|$ orbits stay in similar orbits as the cluster for a relatively long time since the influence of spiral arms is weaker and GMC encounters only happen close to the Galactic plane.    

The average stellar escape rate between GMC encounters for clusters A, B and C ranges between 4.5 to 6.4 stars per Myr which is quite different from what \citet{Hurley2005} found in their $N$-body simulation of M67. Their average escape rate is $\sim 8$ stars per Myr for a cluster with 36000 initial stars with a binary fraction of 50 \% and $r_h = 3.9$ pc. Their cluster is placed at a Galactocentric distance of 8.0 kpc in a static galactic potential which makes our representation of M67's galactic enviroment more realistic. The discrepancy between our clusters and those of \citet{Hurley2005} is dominated by our lack of primordal binaries. This makes the relaxation time for \citet{Hurley2005}'s cluster shorter, leading to a higher escape rate. It will also make the cluster dynamically older and is why in terms of remaining stars after 4 Gyr our clusters are somewhat similar with their cluster having 3520 stars left (870 singles and 1325 binaries). Our galactic enviroment and the GMC encounters for cluster A, B and C are in this way compensating for the lack of primordal binarity in our clusters in terms of ending up with the similar number of stars as \citet{Hurley2005}. If we would want to have a more realistic dynamical version of M67 and add a binary fraction of 50 \% we would need to add an additional number of $\sim10^5$ stars to our initial clusters since the shorter relaxation time would increase the average stellar escape rate. A more massive birth cluster representing M67 could possibly explain the difference in the observed number of around 30 blue stragglers \citep{Deng1999} versus the 20 found in the $N$-body simulations of M67 by \citet{Hurley2005}. A more massive cluster would produce more blue stragglers and mass segregation in the cluster would cause the blue stragglers to sink to the cluster centre where they will be less affected by the galactic tidal forces. This would inadvertently increase the fraction of blue stragglers in the cluster as lower mass stars are tidally stripped from the cluster. We chose not to use binaries in our $N$-body simulations of M67 as a consequnce of computational issues with the binaries not working well with the tidal tensor of \textsc{nbody6tt}. In turn for using \textsc{nbody6tt} and no binaries we get a better representation of the galactic tidal field affecting the clusters, while only lowering the stellar escape rate which should not have significant effect on $f_{\odot}$.

\section{Conclusions}
\label{sec:Conclusions}
In this paper we have simulated the evolution of M67-like clusters in a galactic Milky Way potential, followed the galactic orbits of escaping stars and compared them to the nature of the current solar orbit. We find that the simulated M67 candidates can be divided up into three groups; hot clusters, depleted clusters and scattered clusters. The hot clusters are born with $z$-velocities from the tails of the initial distribution. They therefore have an easier time reaching high $|z|$ without having to go through as many GMC encounters. Depleted clusters and scattered clusters all start out with a low $|v_{z,0}|$ and therefore need more heating by GMC encounters to reach the present day orbit of M67. Depleted clusters go through the largest number of GMC encounters, between 8 and 10, and as a result are depleted of stars to the point where they do not represent M67 at the present day.

All model clusters have stellar escapers that end up on solar-like orbits. Hot clusters have the lowest fraction of escapers that end up on solar-like orbits ($f_{\odot} = 0.06$ \%) and depleted clusters have the highest with $f_{\odot} = 6.61$ \%. Scattered clusters lie in between with an average $f_{\odot}$ of 2.1 \%. 

Hot clusters are similar to the clusters simulated by \citet{Pichardo2012} and we show that it is in fact possible to link the dynamics of these clusters to escapers with solar-like orbits by including GMCs to our galactic model. This dynamical link does not require the stars to escape the cluster with high escape speeds as seen in Figure \ref{fig:v_esc}, with the majority of stars having escape speeds less than 10 \kms. These escape speeds are compatible with the survival of the wide planetary objects in the Solar System for all types of clusters. Stars above $\sim 10$ \kms \, are either black holes or neutron stars that have received kicks. The depleted clusters and scattered clusters correspond to the clusters that are investigated by \citet{Gustafsson2016} where the nature of scatterings between GMC and clusters was explored.  

We have shown that the orbits of stellar escapers in a time-varying galactic potential diffuse from their original values over a few galactic orbits which means that only recent stellar escapers will be able to be dynamically linked to their host cluster. 

By using $f_{\odot}$ from the depleted clusters and scattered clusters we make a rough estimate and calculate that clusters that are destroyed in the Galactic disc have a specific frequency of escapers that end up on solar-like orbits that is $\sim$ 2 times that of escapers from clusters that survive their journey.

\section*{Acknowledgements}
We would like to thank Bengt Gustafsson, Melvyn Davies, Florent Renaud, Guido Moyano and Lennart Lindgren for valuable discussions and suggestions that helped form this paper. This work has been supported by grant 2017-04217 from the Swedish Research Council and the project grant 2014.0017 `IMPACT' from the Knut and Alice Wallenberg Foundation. The simulations were performed on resources provided by the Swedish National Infrastructure for Computing (SNIC) at Lunarc, which we can contribute thanks to grants from The Royal Physiographic Society of Lund.



\bibliographystyle{mnras}
\bibliography{references} 




\bsp	
\label{lastpage}
\end{document}